\documentclass[]{llncs}

\usepackage{amscd,amsmath}
\usepackage{amsfonts}
\usepackage{amssymb}
\usepackage{mathtools}
\usepackage{latexsym}
\usepackage{url}
\usepackage[dvipsnames]{xcolor}
\usepackage{comment}
\usepackage{graphicx}
\usepackage{algorithm,float}
\usepackage{algpseudocode}
\usepackage{braket}
\usepackage[braket, qm]{qcircuit}
\usepackage{etex}
\usepackage{caption} 
\usepackage{tikz}
\usetikzlibrary{arrows, chains, calc, matrix, positioning, scopes, patterns, shapes, fit, decorations.pathreplacing, backgrounds}

\newcommand{\eps}{\varepsilon}

\newcommand{\mR}{\mathbb{R}}
\newcommand{\mZ}{\mathbb{Z}}

\newcommand{\uSVP}{\mathrm{\textsc{u}SVP}}
\newcommand{\BDD}{\mathrm{BDD}}
\newcommand{\LWE}{\mathrm{LWE}}
\newcommand{\DLWE}{\mathrm{dLWE}}

\newcommand{\DCP}{\mathrm{DCP}}

\newcommand{\LMSS}{\mathrm{EDCP}}
\newcommand{\DLMSS}{\mathrm{dEDCP}}
\newcommand{\ULMSS}{\mathrm{U\textnormal{-}EDCP}}

\newcommand{\GLMSS}{\mathrm{G\textnormal{-}EDCP}}
\newcommand{\DGLMSS}{\mathrm{dG\textnormal{-}EDCP}}
\newcommand{\GapSVP}{\mathrm{GapSVP}}

\newcommand{\poly}{\mathrm{poly}}

\newcommand{\supp}{\mathrm{supp}}

\newcommand{\gauss}{\mathcal{D}} 

\renewcommand{\vec}[1]{\mathbf #1}

\newcommand{\Z}{{\mathbb Z}}

\newcommand{\R}{{\mathbb R}}

\newcommand{\hide}[1]{}

\newcommand{\abs}[1]{\lvert#1\rvert}

\newcommand*{\ScProd}[2]{\ensuremath{\langle#1\!\mathbin{,}\!#2\rangle}} 

\newcommand{\bigO}{\mathcal{O}}

\newcommand*{\WLandau}{\varOmega}

\newcommand*{\TLandau}{\varTheta}

\renewcommand{\th}{^{\mathrm{th}}}

\newcommand{\qLat}{\ensuremath{\Lambda_q}}
\newcommand{\qLatp}{\ensuremath{\Lambda_q^{\perp}}}

\newcommand{\Lat}{\ensuremath{\Lambda}}

\DeclareMathSymbol{\mlq}{\mathord}{operators}{``}
\DeclareMathSymbol{\mrq}{\mathord}{operators}{`'}

\newcommand*\widebar[1]{%
	\hbox{%
		\kern0.2em 
		\vbox{%
			\hrule height 0.5pt 
			\kern0.5ex
			\hbox{%
				\kern-0.2em
				\ensuremath{#1}%
				\kern-0.1em
			}%
		}%
	}%
}

\usepackage[normalem]{ulem}
\newcommand{\revdel}[1]{\sout{#1}}
\newcommand{\revadd}[1]{{\color{red}#1}}
\newenvironment{revaddblock}{\color{black}}{}

\begin{document}

\title{Learning With Errors and Extrapolated Dihedral Cosets}
\author{
Zvika Brakerski\thanks{Weizmann Institute of Science, \texttt{zvika.brakerski@weizmann.ac.il}. Supported by the Israel Science Foundation (Grant No. 468/14) and Binational Science Foundation (Grants No. 2016726, 2014276) and ERC Project 756482 REACT.}
\and
Elena Kirshanova\thanks{ENS de Lyon and Laboratoire LIP (U.\ Lyon, CNRS, ENS de Lyon, INRIA, UCBL),
\texttt{elena.kirshanova@ens-lyon.fr}. Supported by ERC Starting Grant ERC-2013-StG-335086-LATTAC.}
\and
Damien Stehl\'e\thanks{ENS de Lyon and Laboratoire LIP (U.\ Lyon, CNRS, ENS de Lyon, INRIA, UCBL),
	\texttt{damien.stehle@ens-lyon.fr}. Supported by ERC Starting Grant ERC-2013-StG-335086-LATTAC.}
\and
Weiqiang Wen\thanks{ENS de Lyon and Laboratoire LIP (U.\ Lyon, CNRS, ENS de Lyon, INRIA, UCBL),
\texttt{weiqiang.wen@ens-lyon.fr}. Supported by ERC Starting Grant ERC-2013-StG-335086-LATTAC.}
}
\institute{}
\date{\today}

\maketitle

\def\dcp{\mathrm{DCP}}
\def\lwe{\mathrm{LWE}}

\begin{abstract}
The hardness of the learning with errors (LWE) problem is one of the most fruitful resources of modern cryptography. In particular, it is one of the most prominent candidates for secure post-quantum cryptography. Understanding its quantum complexity is therefore an important goal.

We show that under quantum polynomial time reductions, LWE is equivalent to a relaxed version of the dihedral coset problem (DCP), which we call extrapolated DCP (eDCP). The extent of extrapolation varies with the LWE noise rate.
By considering different extents of extrapolation, our result generalizes Regev's famous proof that if DCP is in BQP (quantum poly-time) then so is LWE (FOCS~02). We also discuss a connection between eDCP and Childs and Van Dam's algorithm for generalized hidden shift problems (SODA~07).

Our result implies that a BQP solution for LWE might not require the full power of solving DCP, but rather only a solution for its relaxed version, eDCP, which could be easier.
%
%
\end{abstract}

\begin{revaddblock}
\begin{footnotesize}
\sloppy
\noindent
{\bf Revision (v3, November 2026).}
This version corrects some statements of the previous versions (arXiv v1, October~2017, and v2, May~2019). The failure of Lemma~\ref{lem:l2bound} for $m=n$ was observed in~\cite{Fer26}, which also gives a corrected statement and proof for prime~$q$~\cite[Lemma~4.9]{Fer26} (the statement given here holds for any~$q$), as well as the conditions $2\sqrt{2\pi e} \le q^{(m-n)/m} \le 3\sqrt{2\pi e}\, q/\sqrt{m}$ now required in Theorem~\ref{thm:LWEvDCP_ball}~\cite[Section~4.4, conditions (B2)--(B3)]{Fer26}. The other issues were identified, and the corrections were prepared, with the assistance of Claude Opus~5.5 (Anthropic); all corrections were checked by the authors.
\begin{itemize}
\item[$\bullet$] Lemma~\ref{lem:infbound}: the lower bound on $\lambda_1^{\infty}(\Lat_q(\vec{A}))$ is now $q^{(m-n)/m}/4$ instead of $q^{(m-n)/m}/2$, and it holds with probability at least $1-(1+\ln q)(3/4)^m$ instead of $1-2^{-m}$. A proof is given in Appendix~\ref{pro:infbound-v3}.
\item[$\bullet$] Lemma~\ref{lem:l2bound}: it now requires $m \ge 2n$ and $q^{(m-n)/m} \ge 2\sqrt{2\pi e}$, and holds with probability at least $1-\tau(q)(3/4)^m-3(2m+1)^{15}2^{-m/2}$ instead of $1-2^{-m}$, where $\tau(q) = q^{o(1)}$ is the number of divisors of~$q$. A proof is given in Appendix~\ref{pro:l2bound-v3}.
\item[$\bullet$] Lemma~\ref{lem:seperate_gaussian_ball}: it is now stated for any matrix $\vec{A}$ such that $\vec{x} \mapsto \vec{A}\vec{x} \bmod q$ is injective, instead of a uniformly random matrix~$\vec{A}$.
\item[$\bullet$] Theorem~\ref{thm:LWEvDCP}: it now requires $q \ge 3$, and the upper bound on $r$ is now $1/(128m\kappa\alpha\ell q^{n/m})$ instead of $1/(32m\kappa\alpha\ell q^{n/m})$.
\item[$\bullet$] Theorem~\ref{thm:LWEvDCP_ball}: it now requires $m \ge 2n$ and $2\sqrt{2\pi e} \le q^{(m-n)/m} \le 3\sqrt{2\pi e}\, q/\sqrt{m}$, and $m = \Omega(\kappa+\log q)$ instead of $m = \Omega(\kappa)$.
\item[$\bullet$] Corollary~\ref{cor:dLWEvDCP}: it now requires $m \ge 2(n+1)$ and $2\sqrt{2\pi e} \le q^{(m-n-1)/m} \le 3\sqrt{2\pi e}\, q/\sqrt{m}$, and $m = \Omega(\kappa+\log q)$ instead of $m = \Omega(\kappa)$. The upper bound on~$r$ now includes a factor~$\varepsilon$, the advantage of the $\DGLMSS$ oracle.
\item[$\bullet$] Theorem~\ref{thm:vDCPLWE2}: it now explicitly requires $N > 2\sqrt{\kappa}\, r$. Its proof now accounts for the fact that the measurement produces errors of parameter $N/(\sqrt{2}r)$ rather than $N/r$.
\item[$\bullet$] Corollary~\ref{thm:dvDCPLWE}: it now explicitly requires $r = \Omega(\sqrt{\kappa})$ and $N > 2\sqrt{\kappa}\, r$.
\end{itemize}
\end{footnotesize}
\end{revaddblock}



\section{Introduction}
\label{sec:intro}

The Learning With Errors problem LWE$_{n,q,\alpha}$ with parameters 
$n,q \in \mZ$ and $\alpha \in (0,1)$
 consists in finding a vector~$\vec{s} \in \mZ_q^n$ from arbitrarily 
many samples $(\vec{a}_i, \langle \vec{a}_i, \vec{s}\rangle + e_i) 
\in \mZ_q^n \times \mZ_q$, 
where $\vec{a}_i$ is uniformly sampled in~$\mZ_q^n$ and $e_i$ is sampled 
from~$\gauss_{\mZ, \alpha q}$, the discrete Gaussian distribution of standard 
deviation parameter~$\alpha q$ (i.e., the distribution such that $\gauss_{\mZ, \alpha q}(k) \sim \exp(-\pi k^2/(\alpha q)^2)$ for all $k \in\mZ$).
Since its introduction by Regev~\cite{Regev05,Regev09}, LWE has served as a security
foundation for numerous cryptographic primitives (see e.g.\ an overview in \cite{Peikert16}). The cryptographic attractiveness of 
LWE stems from two particularly desirable properties. First, its algebraic 
simplicity enables the 
design of primitives with advanced functionalities, such as fully homomorphic 
encryption~\cite{BrVa11a}, attribute-based encryption for all circuits~\cite{GVW13} 
and (single key) functional encryption~\cite{GKPVZ13}. Second, LWE is conjectured
hard even in the context of quantum computations, making it one of the most appealing candidate
security foundations for post-quantum cryptography~\cite{BeBuDa08}.
Current quantum algorithms for LWE do not outperform classical ones, but it is not clear whether this is inherent (for example, it is known that LWE is \emph{easier} than the Dihedral Coset Problem under polynomial-time reductions, see below). In this work, we characterize the quantum hardness of LWE under polynomial-time reductions and show that it
%
%
%
is computationally equivalent (up to small parameter losses)
to a quantum problem closely related to the aforementioned Dihedral Coset Problem.

\paragraph{$\LWE$, Lattices and the Dihedral Coset Problem.}
LWE is tightly connected to worst-case approximation problems over Euclidean lattices. In particular, LWE is an (average-case) instance of the Bounded Distance Decoding problem (BDD) (see, e.g., \cite[Section~5.4]{MiRe09}), but is also known to be as hard as \emph{worst-case} BDD (with some polynomial loss in parameters)~\cite{Regev09}. BDD is the problem of finding the closest lattice vector to a given target point which is promised to be very close to the lattice (formally, closer than $\lambda_1/\gamma$ where $\lambda_1$ is the length of the shortest non-zero vector). Classical and quantum connections between BDD and other problems such as SIVP, GapSVP, uSVP are also known \cite{BLPRS13,LyMi09,Peikert09,Regev09}.

%

Regev~\cite{Regev02,Regev04b} showed that uSVP, and therefore also BDD and LWE, are no harder to solve than the quantumly-defined Dihedral Coset Problem (DCP). An instance of DCP$_{N, \ell}$, for integer parameters~$N$ and~$\ell$, consists of~$\ell$ quantum registers in superposition $\ket{0,x_k} + \ket{1, x_k+s}$, with a common $s \in \mZ_N$ and random and independent $x_k\in \mZ_N$ for $k \in [\ell]$. The goal is to find $s$ (information theoretically $\ell = \bigO(\log N)$ is sufficient for this task \cite{EtHo99}). We note that Regev considered a variant with unbounded number of registers, but where a fraction of them is faulty (a faulty state is of the form $\ket{b, x_k}$ for arbitrary $b \in \{0,1\}, x_k \in \Z_N$). In our work, we assume a non-faulty formulation of DCP. 

Still, it is quite possible that DCP is in fact much harder to solve than LWE. The best known algorithm for DCP, due to Kuperberg~\cite{Kuperberg05}, runs in time $2^{\bigO(\log\ell + \log N/ \log\ell)}$ which does not improve upon classical methods for solving LWE. Other variants of the problem were explored in \cite{EtHo99,FIMSS14}, and of particular relevance to this work is a ``vector'' variant of the problem where $\mZ_N$ is replaced with $\mZ^n_{q}$ (i.e.\ $s$ and $x_k$ are now vectors). These problems behave similarly to DCP with $N=q^n$.

Finally, Regev showed that DCP can be solved given efficient algorithms for the subset-sum problem (which is classically defined), however in a regime of parameters that appears harder to solve than LWE itself.

\paragraph{Extrapolated DCP.} The focus of this work is a generalization of the DCP problem, i.e.\ rather than considering registers containing $\ket{0,x_k} + \ket{1, x_k+s}$, we allow (1) $x_k$'s and $s$ be $n$-dimensional vectors, and (2) other than non-uniform distribution for amplitudes. 
We name this problem Extrapolated DCP ($\LMSS$) as its input registers have more extrapolated states. 
To be more precise, $\LMSS_{n,N,f}^{\ell}$, with parameters three integers~$n,N,\ell$ and a function~$f: \mZ \mapsto \mathbb{C}$ with~$\sum_{j\in \mZ} j \cdot |f(j)|^2 < +\infty$,  
consists in recovering $\vec{s}\in\mZ_N^n$ from 
the following $\ell$ states over~$\mZ \times \mZ_N^n$:
\[
\left\{ \frac{1}{\sqrt{\sum_{j \in \mZ} |f(j)|^2}}  \cdot 
\sum_{j \in \mZ} f(j) \ket{j,\vec{x}_k + j \cdot \vec{s}} \right\}_{k \leq \ell}, 
\]  
where the $\vec{x}_k$'s are arbitrary in~$\mZ_N^n$.\footnote{Note that the  assumption on~$f$  
implies, via Markov's inequality, that one may restrict the sum to a finite 
index set and obtain a superposition which remains within negligible $\ell_2$ distance 
from the  countable superposition above.} Note that DCP is the special case  of $\LMSS$ for $n=1$ and $f$ being the indicator function of $\{0, 1\}$.


In~\cite{ChDa07}, Childs and van Dam consider a special 
case of $\LMSS$ where~$f$ is the indicator 
function of~$\{0,\ldots,M-1\}$ for some integer~$M$, which we will refer to as 
uniform $\LMSS$ (or, $\ULMSS_{n,N,M}^{\ell}$). 
\paragraph{Our Main Result.}
We show that up to polynomial loss in parameters,  $\ULMSS$  is equivalent to LWE. Thus we provide a formulation of the hardness assumption underlying lattice-based cryptography in terms of the (generalized) Dihedral Coset Problem.

\begin{theorem}[Informal]
	\label{th:main}
	There exists a quantum polynomial-time reduction from $\LWE_{n,q,\alpha}$ 
	to $\ULMSS_{n,N,M}^{\ell}$, with~$N = q$,
	$\ell = \poly(n\log q)$ and $M = \frac{\poly(n\log q)}{\alpha}$. 
	Conversely,  there exists a polynomial-time reduction from $\ULMSS$  to $\LWE$ with the same parameter relationships, up 
	to $\poly(n\log q)$ factors.
\end{theorem} 

Our proof crucially relies on a special case of $\LMSS$ where $f$ is a Gaussian weight function with standard deviation parameter $r$. We call this problem Gaussian $\LMSS$ ($\GLMSS$). We show that $\GLMSS$ and $\ULMSS$ are equivalent up to small parameter losses.

$\LMSS$ is analogous to $\LWE$ in many aspects. The decisional version of $\LWE$ ($\DLWE$) asks to distinguish between $\LWE$ samples and random samples of the form $(\vec{a},b)\in \mZ_q^n\times \mZ_q$ where both components are chosen uniformly at random. Similarly, we also consider the decisional version of $\LMSS$, denoted by $\DLMSS$. In $\DLMSS_{n,N,f}$, we are asked to distinguish between an $\LMSS$ state and a state of the form
\[
\Ket{j}\Ket{\vec{x} \bmod N},
\]
where $j$ is distributed according to the function $\abs{f}^2$,
and $\vec{x}\in \mZ_N^n$ is uniformly chosen. $\LMSS$ enjoys a reduction between its search and decisional variants via $\LWE$.


\paragraph{Related work.}
In~\cite{ChDa07}, Childs and van Dam show that $\ULMSS_{1, N, M}^{\ell}$ reduces to the problem 
of finding all the solutions $\vec{b} \in \{0,\ldots,M-1\}^k$ to the 
equation~$\langle \vec{b},\vec{x} \rangle = w \bmod N$, where $\vec{x}$ and $w$ 
are given and uniformly random modulo~$N$. They interpret this as an
integer linear program and use lattice reduction, within Lenstra's 
algorithm~\cite{Lenstra83}, to solve it. This leads to a  polynomial-time 
algorithm for $\ULMSS_{1, N, M}^{\ell}$ when $M = \lfloor N^{1/k} \rfloor$ and $\ell \geq k$, for any~$k\geq 3$. 
Interestingly, finding small solutions to the 
equation~$\langle \vec{b},\vec{x} \rangle = w \bmod N$ is a special case of the 
Inhomogeneous Small Integer Solution problem~\cite{GePeVa08} (ISIS), which consists in 
finding a small-norm~$\vec{x}$ such that $\vec{B} \vec{x} = \vec{w} \bmod q$, 
with~$\vec{B} \in \mZ_q^{n \times m}$ and $\vec{w} \in \mZ_q^n$ 
uniform (where~$q,n,m$ are integer parameters). A reduction from the 
homogeneous SIS (i.e., with~$\vec{w} = \vec{0}$ and $\vec{x}\neq \vec{0}$) to LWE was provided 
in~\cite{SSTX09}. It does not seem possible to derive from it a reduction 
from $\LMSS$ to LWE via the Childs and van Dam variant of ISIS, most notably because
the reduction from~\cite{SSTX09}  does not provide a way to compute all 
ISIS solutions within a  box~$\{0,1,\ldots,M-1\}^k$.

It is not hard to see that, at least so long as $M$ is polynomial, a solution to DCP implies a solution to $\LMSS_{n,N,M}^{\ell}$. Therefore our result implies \cite{Regev02} as a special case. On the other extreme, our result also subsumes \cite{ChDa07} since the LLL algorithm~\cite{LeLeLo82} 
can be used to solve $\LWE_{n,q,\alpha}$ in polynomial time when~$1/\alpha$ and~$q$ are~$2^{\Theta(n)}$, which implies a polynomial-time algorithm for 
$\LMSS$ for $M = 2^{\Theta(\sqrt{n\log N})}$, significantly improving Childs and van Dam's $M= 2^{\eps n\log N}$.

Finally,
we observe that the $\LWE$ to $\ULMSS$ reduction (and the uSVP to DCP
reduction from~\cite{Regev04b}) can be adapted to a uSVP to $\ULMSS$ reduction, 
as explained below. Combining this adaptation with the reduction from $\ULMSS$ 
to LWE (via $\GLMSS$) provides a novel quantum reduction from worst-case 
lattice problems to $\LWE$. However,
it does not seem to have advantages compared to~\cite{Regev09}.

\subsection{Technical overview} \label{subsect:TechnicalOverview}

As mentioned above, the hardness of $\LWE$ is essentially invariant so long as $n \log q$ is preserved, and therefore we restrict our attention in this overview to the one-dimensional setting. A crucial ingredient in our reduction is a weighted version of $\LMSS$, denoted by $\GLMSS$ and quantified by 
a Gaussian weight function $f_r(j) =~\mkern-10mu \rho_r(j)=\exp(-\pi j^2/r^2)$, for some standard deviation parameter~$r$. We refer to this problem as Gaussian $\LMSS$ ($\GLMSS$).


%
%
\def\w{5.3} 
\def\h{3} 
\def\vd{1.5} 
\def\hd{1.5} 

\renewcommand{\longleftrightarrow}{\hspace{6pt} \tikz[baseline=3.5ex] \draw[<->] (0,0.7) -- (1.1,0.7); \hspace{6pt}}
\renewcommand{\longrightarrow}{\hspace{5pt} \tikz[baseline=-0.5ex] \draw[->] (0,0.0) -- (0.8,0.0); \hspace{5pt}}

\begin{figure}
	\centering
	\begin{tikzpicture}
		\coordinate (ll) at (-2, -2); 
		\draw[black, thin, rounded corners=10pt] (ll) rectangle ($(ll)+(\w, \h)$) node[pos=.5] (LWE) {
			\large
			$ \LWE_{n, \tfrac{1}{\alpha},  \alpha}$
			};
		
		\draw[black, thin, name=LMSS, rounded corners=10pt] ($(ll)-(0,\vd+\h)$) rectangle ($(ll)+(\w, -\vd)$){}; 
		node[matrix] (LMSS) {
			\node (LMSS1) at (0.7,-4.2) {$\ULMSS_{ \big(\tfrac{1}{\alpha}\big)^n\mkern-10mu, \poly(n), \tfrac{1}{\alpha}}$};
			\node (LMSS2) at ($(LMSS1)+(0, -1.5)$) {$\GLMSS_{\big(\tfrac{1}{\alpha}\big)^n\mkern-10mu, \poly(n), \frac{1}{\alpha}}$};
			
			\draw[->] ($(LMSS1)+(-.3, -.4)$)--($(LMSS2)+(-.3, .4)$) node[midway, xshift=-2em]{\scriptsize Lemma~\ref{lem:g2u}};
			\draw[<-] ($(LMSS1)+(.3, -.4)$)--($(LMSS2)+(.3, .4)$) node[midway, xshift=2em]{\scriptsize Lemma~\ref{lem:u2g}};
		};
		
		\draw[black, thin, rounded corners=10pt] ($(ll)+(\w+\hd,0)$) rectangle ($(ll)+(2*\w+\hd,\h)$){};
		node[matrix]{
			\node (BDD) at (6.0,0.2) {$\BDD_{\alpha}$};
			\node (uSVP) at ($(BDD)+(3, 0)$) {$\uSVP_{1/\alpha}$};
			\node (GapSVP) at (7.7, -1.2) {$\GapSVP_{1/\alpha}$};
			\draw[<->] (BDD)--(uSVP);
			\draw[<->] (uSVP)--(GapSVP);
			\draw[<->] (GapSVP)--(BDD);
		};
		
		\draw[black, thin, name=Reg, rounded corners=10pt]  ($(ll)+(\w+\hd,-\vd-\h)$) rectangle ($(ll)+(2*\w+\hd, -\vd)$) node[pos=.5] (Reg) {$\DCP_{\big(\tfrac{1}{\alpha}\big)^n\mkern-10mu, \poly(n)} \longrightarrow$ Subset Sum};
		
		
		\draw[->] ($(ll)+(\w/2-.3,-0.1)$)--($(ll)+(\w/2-.3, -\vd+0.1)$) node[midway, xshift=-4em]{Theorems~\ref{thm:LWEvDCP}, \ref{thm:LWEvDCP_ball}}; 
		
		\draw[<-] ($(ll)+(\w/2+.3,-0.1)$)--($(ll)+(\w/2+.3, -\vd+0.1)$)node[midway, xshift=3em]{Theorem~\ref{thm:vDCPLWE2}}; 
		
		\draw[<->] ($(ll)+(\w+0.1,\h/2)$)--($(ll)+(\w+\hd-0.1, \h/2)$) node[above, midway]{\cite{Regev05,Regev09}}; 
		
		\draw[->]
		($(ll)+(\w+0.1, -\hd-\h/2+.2)$) -- ($(ll)+(\w+\hd-0.1, -\hd-\h/2+.2)$)
		node[above, midway]{\scriptsize Lemma~\ref{lem:vDCP2DCP_prime}}; 
		
		\draw[->]
		($(ll)+(\w+\hd+.7,-.1)$) -- ($(ll)+(\w+\hd+.7, -\hd-\h/2+0.6)$) node[midway, right, yshift=1em]{\cite{Regev02}}; 
		
		
		\node[] at (7.6, -0.3) {\cite{LyMi09}};
		\node[] at (7.8, -4.7) {\cite{Regev02}};
		
		
	\end{tikzpicture}%
	\caption{Graph of reductions between the $\LWE$ problem (upper-left), worst-case lattice problems (upper-right), combinatorial problems (lower-right) and the Extrapolated Dihedral Coset problems (lower-left). Parameters $\alpha$ are given up to $\poly(n)$-factors, where $n$ is the dimension of the  $\LWE$ problem. The same $n$ stands for the lattice-dimension considered in problems of the upper-right corner. The subset-sum problem stated in the lower-right corner is of  density $\approx 1$ (in particular, the expected number of solutions is constant). }
	\label{fig:GraphReductionI}
\end{figure}


\paragraph{Reducing $\GLMSS$ to $\LWE$.}
Given an $\GLMSS$ state as input, our reduction efficiently transforms it into a classical $\LWE$ sample with constant success probability. Thus, making only one query to the $\LWE$ oracle, we are able to solve $\GLMSS$.
More precisely, the reduction input consists of a normalized state corresponding 
to~$\sum_{j\in\mZ_N} \rho_r(j)\ket{j} \ket{x + j \cdot s \bmod N}$, for some integers $r \ll N$. One can think of $N$ as  the $\LWE$ modulus and of $r$ as the standard deviation parameter of the $\LWE$ error.

Our first step is to apply a quantum Fourier transform over $\mZ_N$ to the second register. This gives us a quantum superposition of the form:
\begin{equation*}
\sum_{a\in \mZ_N} 
\sum_{j\in \mZ_N} \omega_N^{a\cdot (x+j\cdot s)}\cdot 
\rho_{r}(j)\Ket{j} \Ket{a}.
\end{equation*}
where~$\omega_N = \exp(2i\pi / N)$.
We then measure the second register and obtain a value~$\widehat{a} \in \mZ_N$. This leaves us with the state:
\begin{equation*}
\sum_{j\in \mZ_N} \omega_N^{j\cdot \widehat{a}\cdot s}\cdot \rho_{r}(j) \Ket{j}\Ket{\widehat{a}}.
\end{equation*}
Note that $\widehat{a}$ is uniformly random over $\mZ_N$, which at the end serves as the first component of LWE sample. The exponent of relative phase in current state has a form similar to the second component of LWE sample but without noise. Now we can benefit from the first register, which stores a superposition corresponding to a Gaussian distribution over $\mZ_N$ with standard deviation $r$. Applying a second quantum Fourier transform over~$\mZ_N$ to the first register gives us a quantum superposition of the form:
\begin{equation*}
\sum_{b\in \mZ_N}\sum_{j\in \mZ_N} \omega_N^{j\cdot (\widehat{a}\cdot s + b)}\cdot \rho_{r}(j) \Ket{b}.
\end{equation*}

Now the second component of the $\LWE$ sample $\widehat{a}\cdot s + b$ is stored in the phase (up to a factor $j$). Omitting the exponentially small Gaussian tail, we assume the summation for $j$ is taken over the integers. An application of the Poisson summation formula transfers $\widehat{a}\cdot s + b$ into a shift of the Gaussian distribution defined over $\mZ$. In other words, the received state is exponentially close to the superposition:
\begin{equation*}
\sum_{e\in \mZ_N}\rho_{1/r}\Big(\frac{e}{N}\Big) \Ket{-\widehat{a}\cdot s + e}.
\end{equation*}
Once we measure the state above, we obtain a value $-\widehat{a}\cdot s +e$, where $e \hookleftarrow \mathcal{D}_{\mZ,\text{\revdel{$N/r$}}\text{\revadd{$N/(\sqrt{2}r)$}}}$. Together with already known $\widehat{a}$, this gives us an $\LWE$ sample:
\[
\left( -\widehat{a}, -\widehat{a}\cdot s +e\right).
\]

In case the input state is of the form $\Ket{j}\Ket{x \bmod N}$, where $j$ is distributed according to the function $\rho_{r}^2$, and $x\in \mZ_N$ is uniformly chosen (the decisional case), the reduction outlined above outputs a uniform random pair $(a,b)$ from $\mZ_N \times \mZ_N$. This gives a reduction from decisional version of $\GLMSS$ to decisional version of $\LWE$.

\paragraph{Reducing $\LWE$ to $\GLMSS$.}
Our reduction from LWE to $\GLMSS$ follows the general design 
of Regev's reduction from uSVP to DCP~\cite{Regev04b},  with several 
twists that enable simplifications and improvements. We note that this
reduction is folklore,\footnote{ \url{https://groups.google.com/d/msg/cryptanalytic-algorithms/uhr6gGrVkIk/XxEv4uvEBwAJ} 
} although we could not find it described explicitly.

First, the use of LWE rather than uSVP allows us to avoid 
Regev's initial sub-reduction 
from uSVP to BDD, as LWE is  a randomized variant of BDD. Indeed, if we 
consider $m$ samples~$(a_i, a_i \cdot s + e_i)$  from LWE$_{n,q,\alpha}$, then we have 
a BDD instance for the lattice~$\Lambda = \vec{A} \mZ_q + q \mZ^m$ and the target 
vector~$\vec{t} = \vec{b} + \vec{e} \in \mZ^m$
with~$\vec{b} \in \Lambda$ satisfying~$\vec{b} = \vec{A}  \cdot s \bmod q$. 

As Regev's, our reduction proceeds by subdividing the ambient space~$\mR^m$ 
with a coarse grid, setting the cell width between~$\|\vec{e}\|$ 
and~$\lambda_1(\Lambda)$. We map each point $\vec{y} \in \mR^m$ to a cell~$\phi(\vec{y})$.
By choice of the cell width, we have~$\phi(\vec{c}_1) \neq \phi(\vec{c}_2)$ for any 
$\vec{c}_1 \neq \vec{c}_2$ in~$\Lambda$.
Also for any~$\vec{c} \in \mR^m$, the vectors~$\vec{c}$ and~$\vec{c}+\vec{e}$ are 
most likely
mapped to the same cell, as $\vec{e}$ is short.  This intuition fails if a border  between two cells falls close to~$\vec{c}$. This 
(rare but non-negligibly so) event is the source of  the limitation on the 
number~$\ell$ of DCP/$\LMSS$ states produced by the reduction. The space subdivision 
by a grid is illustrated in Figure~\ref{fig:grid}.

\usetikzlibrary{arrows.meta}

\pgfdeclareradialshading{ring}{\pgfpoint{0cm}{0cm}}%
{rgb(0cm)=(1,1,1);
rgb(0.7cm)=(1,1,1);
rgb(0.719cm)=(1,1,1);
rgb(0.72cm)=(0.975,0,0);
rgb(0.9cm)=(1,1,1)}

\begin{figure}
\centering
\begin{tikzpicture}[scale=1.8]

    \def\radius{0.14cm}
    \def\corner{0.01cm}

    \def\level{4}
    \foreach \x in {3,5}
    {
      \filldraw[white,even odd rule,inner color=black!80,outer color=black!20,line width=\corner] (\x+1,\level) circle (\radius);
      \draw [fill,black] (\x+1,\level) circle [radius=0.02cm];
    }

    \def\level{5}
    \foreach \x in {3,5,7}
    {
      \filldraw[white,even odd rule,inner color=black!80,outer color=black!20,line width=\corner] (\x,\level) circle (\radius);
      \draw [fill,black] (\x,\level) circle [radius=0.02cm];
    }

    \def\level{6}
    \foreach \x in {3,5}
    {
      \filldraw[white,even odd rule,inner color=black!80,outer color=black!20,line width=\corner] (\x+1,\level) circle (\radius);
      \draw [fill,black] (\x+1,\level) circle [radius=0.02cm];
    }

    \foreach \x in {1,2,3}
    {
      \draw [black] [-,dashed] (0.65*\x+5.5,2.65+0.65*1) -- (0.65*\x+5.5,7.1-0.65*1);
      \draw [black] [-,dashed] (-0.65*\x+5.5,2.65+0.65*1) -- (-0.65*\x+5.5,7.1-0.65*1);
    }
    \draw [black] [-,dashed] (-0.65*4+5.5,2.65+0.65*1) -- (-0.65*4+5.5,7.1-0.65*1);
    \draw [black] [-,dashed] (0+5.5,2.65+0.65*1) -- (0+5.5,7.1-0.65*1);
    \foreach \x in {1}
    {
      \draw [black] [-,dashed] (0.75+0.65*3,0.65*\x+5.5) -- (9.58-0.65*3,0.65*\x+5.5);
      \draw [black] [-,dashed] (0.75+0.65*3,-0.65*\x+5.5) -- (9.58-0.65*3,-0.65*\x+5.5);
    }
    \draw [black] [-,dashed] (0.75+0.65*3,-0.65*2+5.5) -- (9.58-0.65*3,-0.65*2+5.5);
    \draw [black] [-,dashed] (0.75+0.65*3,0+5.5) -- (9.58-0.65*3,0+5.5);
    \draw [black] [-,dashed] (0.75+0.65*3,-0.65*3+5.5) -- (9.58-0.65*3,-0.65*3+5.5);

    \draw [black,line width=0.02cm] (-1.05+5.6,-0.65+5.5) -- (-1.05+5.8, -0.65+5.5);
    \draw [black,line width=0.02cm] (-1.05+5.6,0+5.5) -- (-1.05+5.8, 0+5.5);
    \draw [black,line width=0.02cm, <->,>=stealth] (-1.05+5.7,-0.65+5.5) -- (-1.05+5.7, 0+5.5);
    \draw (-1.45+5.95,-0.325+5.5) node[black,scale=0.6] {\Large $d$};

    \draw [black,line width=0.02cm] (-0.65+5.5,-1.05+5.6) -- (-0.65+5.5,-1.05+5.8);
    \draw [black,line width=0.02cm] (0+5.5,-1.05+5.6) -- (0+5.5,-1.05+5.8);
    \draw [black,line width=0.02cm, <->, >=stealth] (-0.65+5.5,-1.05+5.7) -- (0+5.5,-1.05+5.7);
    \draw (-0.325+5.5,-1.45+5.95) node[black,scale=0.6] {\Large $d$};

    \draw [black,line width=0.02cm, -{Latex[length=0.5mm]}] (5,5) -- (5.10, 5.10);
    \draw (5.25, 5.15) node[scale=0.6]{\Large $r$};
\end{tikzpicture}
\caption{A visualization of the space subdivision. Each radially shaded disk has width $r$,  the upper bound of the error $\|\vec{e}\|$. Each cell has width $d$, chosen to be between $\|\vec{e}\|$ and $\lambda_1(L)/\sqrt{m}$. Note that the grid intersects the left-most disk, potentially leading to an error in the reduction. 
	\label{fig:grid}}
\end{figure}

Regev's reduction and ours differ in the way the grid is used to create the DCP/$\LMSS$
states. Let us first briefly recall the core of Regev's reduction.
Let~$\vec{B} = (\vec{b}_1,\ldots,\vec{b}_m)$ be 
a basis of~$\Lambda$ and subtract an appropriate combination of the $\vec{b}_i$'s 
from~$\vec{t}$ to get~$\vec{t}'$ so that the coordinates~$\vec{x}'$ of the  
closest vector~$\vec{b}' \in \Lambda$ 
to~$\vec{t}'$ with respect to  the~$\vec{b}_i$'s are~$\leq 2^{m}$ (this may be achieved 
using LLL~\cite{LeLeLo82} and Babai's nearest plane algorithm~\cite{Babai86}).
The first step is the creation of a superposition
\begin{align*}
\hspace*{-.1cm}
& \sum_{\substack{\vec{x} \in \mZ^m \\ \|\vec{x}\|_{\infty} \leq 2^{2m}}} 
\hspace*{-.4cm}
\left( \ket{0, \vec{x}, \phi(\vec{B} \vec{x})} +
\ket{1, \vec{x}, \phi(\vec{B} \vec{x}-\vec{t}')} \right)
=  \\
& \ket{0} 
\hspace*{-.4cm}
\sum_{\substack{\vec{x} \in \mZ^m \\ \|\vec{x}\|_{\infty} \leq 2^{2m}}}
\hspace*{-.4cm} 
\ket{\vec{x}, \phi(\vec{B}\vec{x})} +
\ket{1}
\hspace*{-.7cm}
\sum_{\substack{\vec{x} \in \mZ^m\\ \|\vec{x}+\vec{x}'\|_{\infty} \leq 2^{2m}}} 
\hspace*{-.7cm} 
\ket{\vec{x}+\vec{x}', \phi(\vec{B} \vec{x} -\vec{e})},
\end{align*}
where the equality holds by a change of variable. By measuring the last 
register, with overwhelming probability this collapses to 
$\ket{0} \ket{\vec{x}_k} +  \ket{1} \ket{\vec{x}_k +\vec{x}'}$, which 
corresponds to an $m$-dimensional DCP input state 
with modulus~$2^{\bigO(m)}$. The whole process can be repeated multiple times 
using the same input vector~$\vec{t}$, and results in different~$\vec{x}_k$'s
but a common~$\vec{x}'$. Each iteration may fail because of an ill-placed 
cell delimitation, or if $\vec{x}_k +  \vec{x}'$ has a coordinate 
whose magnitude is larger than~$2^{2m}$. This leads to a bounded number
of correct DCP input states. Finally, $m$-dimensional DCP can be reduced to 
1-dimensional DCP, with a significant modulus increase: the resulting
modulus~$N$ is~$2^{\bigO(m^2)}$.

Instead of using a superposition based on  the coordinates  with respect to a basis, we 
exploit the special form of~$\Lambda = \vec{a} \mZ_q +q \mZ^m$ (w.l.o.g., assume 1-dimensional $\LWE$, \cite{BLPRS13}). We start with the 
following superposition:
\[
\sum_{x \in \mZ_q} \mkern-5mu
\ket{0, x, \phi(\vec{a} x)} +
\ket{1, x, \phi(\vec{a} x-\vec{t})}
\mkern-3mu = \mkern-3mu
\ket{0} \mkern-7mu \sum_{x\in \mZ_q}  \mkern-4mu \ket{x, \phi(\vec{a}  x)} +
\ket{1}\mkern-7mu \sum_{x\in \mZ_q}  \mkern-4mu \ket{x +s, \phi(\vec{a} x - \vec{e})}.
\]
We then measure the last register (classically known and omitted) and hopefully obtain a 
superposition~$\ket{0} \ket{x} +  \ket{1} \ket{x + s}$.
This approach has several notable advantages. First, by using a grid over the 
torus~$\mR^m / q \mR^m$,
the only source of failure is the position of the cell delimitation
(coordinates cannot spill over, they wrap around). Second, we directly end up 
with a DCP state, not a vectorial variant thereof. 
Third, and most importantly, the DCP modulus~$N$ is 
only~$q$ and not~$2^{\bigO(m^2)}$. Note that~$m$ should be set as $\Omega(\log q)$ for $s$ to be uniquely determined by the LWE samples. This improvement results in a much tighter reduction.

The improvement stems from the use of a small modulus~$q$ rather than large integer coordinates. It is possible to obtain such a small DCP modulus while starting from BDD (rather than LWE), by modifying 
Regev's reduction as follows. One may first reduce BDD to a variant thereof that asks 
to find the coordinates of the BDD solution modulo a small modulus~$q$ rather than over the integers. Such a reduction is 
presented in~\cite[Lemma~3.5]{Regev09}. One may then reduce this BDD variant to DCP as we proceed for LWE. Note that 
this transformation makes the BDD to DCP reduction from~\cite{Regev04b} iterative: the DCP oracle is called several 
times, and the input of an oracle call depends on the output of the previous oracle calls. This is akin to the phenomenon 
described in the \emph{open questions} paragraph from~\cite{BLPRS13}.

A further difference between our reduction and the one from~\cite{Regev04b} is that we consider larger multiples of~$s$ in the input 
superposition to obtain a state of the form~$\sum_j \rho_r(j)\ket{j} \ket{x +j s}$,
with~$r\approx 1/\alpha$ (up to polynomial factors). This does not lead to any extra complication, but leads 
us to $\GLMSS$ rather than DCP, which we crucially need to allow for 
a converse reduction. We conjecture 
that $\GLMSS$ is strictly easier than DCP.

As Regev~\cite{Regev04}, we can also improve the resulting deviation parameter $r$ of $\GLMSS$ by a factor of $\sqrt{m}$ using balls' intersections rather than cube separation. We consider intersections of balls drawn around $\vec{a}\cdot s$ and its noisy shifts. The radius $R$ of each ball is set to be the largest value such that the balls arising from different $s$ (and their shifts) do not intersect. We are interested in the intersection area of the balls drawn around $\pm s, \pm 2s$, etc. Following Regev~\cite{Regev04}, this area is large enough to guarantee that once we measure, we hit a point from the intersection of all the balls (see grey areas in Figure~\ref{fig:ball}).

\begin{figure}
\centering
\begin{tikzpicture}[scale=1.8]

    \def\radius{0.60cm}
    \def\corner{0.005cm}

    \def\level{4}
    \foreach \x in {3,5}
    {
      \foreach \y in {1}
      {
        \begin{scope}
        \clip (\x+1+0.05*\y,\level+0.05*\y) circle (\radius);
        \filldraw[gray!40,line width=\corner] (\x+1-0.05,\level-0.05) circle (\radius);
        \end{scope}
      }
      \foreach \y in {-1,1}
      {
        \draw[red!65,line width=\corner] (\x+1+0.05*\y,\level+0.05*\y) circle (\radius);
        \draw [fill,red!65] (\x+1+0.05*\y,\level+0.05*\y) circle [radius=0.005cm];
      }
      \foreach \y in {0}
      {
        \draw[black,line width=\corner] (\x+1+0.05*\y,\level+0.05*\y) circle (\radius);
        \draw [fill,black] (\x+1+0.05*\y,\level+0.05*\y) circle [radius=0.005cm];
      }
    }

    \def\level{5}
    \foreach \x in {3,5,7}
    {
      \foreach \y in {1}
      {
        \begin{scope}
        \clip (\x+0.05*\y,\level+0.05*\y) circle (\radius);
        \filldraw[gray!40,line width=\corner] (\x-0.05,\level-0.05) circle (\radius);
        \end{scope}
      }
      \foreach \y in {-1,1}
      {
        \draw[red!65,line width=\corner] (\x+0.05*\y,\level+0.05*\y) circle (\radius);
        \draw [fill,red!65] (\x+0.05*\y,\level+0.05*\y) circle [radius=0.005cm];
      }
      \foreach \y in {0}
      {
        \draw[black,line width=\corner] (\x+0.05*\y,\level+0.05*\y) circle (\radius);
        \draw [fill,black] (\x+0.05*\y,\level+0.05*\y) circle [radius=0.005cm];
      }
    }

    \def\level{6}
    \foreach \x in {3,5}
    {
      \foreach \y in {1}
      {
        \begin{scope}
        \clip (\x+1+0.05*\y,\level+0.05*\y) circle (\radius);
        \filldraw[gray!40,line width=\corner] (\x+1-0.05,\level-0.05) circle (\radius);
        \end{scope}
      }
      \foreach \y in {-1,1}
      {
        \draw[red!65,line width=\corner] (\x+1+0.05*\y,\level+0.05*\y) circle (\radius);
        \draw [fill,red!65] (\x+1+0.05*\y,\level+0.05*\y) circle [radius=0.005cm];
      }
      \foreach \y in {0}
      {
        \draw[black,line width=\corner] (\x+1+0.05*\y,\level+0.05*\y) circle (\radius);
        \draw [fill,black] (\x+1+0.05*\y,\level+0.05*\y) circle [radius=0.005cm];
      }
    }

    \draw [black,line width=0.008cm, <->,>=stealth] (4+0.05*0,6+0.05*0) -- (5+0.05*0,5+0.05*0);
    \draw (4.5,5.4) node[black,scale=0.5,anchor=north east] {\Large $\lambda_1(\Lambda_q(\textbf{a}))$};
    \draw [black,line width=0.01cm ] (5-0.05,5-0.05) -- (5+0.05,5+0.05);
   \draw [line width=0.01cm, decorate,decoration={brace,amplitude=1pt,mirror,raise=0.02cm}]
   (5-0.05,5-0.05) --(5+0.05,5+0.05) node [black,midway,xshift=0.15cm, yshift=-0.1cm] {\scriptsize $d$};

    \draw [red!65,line width=0.008cm, ->,>=stealth] (5+0.05,5+0.05) -- (5+0.05+0.1,5+0.05+0.60);
    \draw (5.10,5.30) node[black,scale=0.5,anchor=west] {\Large $R$};

\end{tikzpicture}
	\caption{A visualization of the balls' intersections. The lattice points (black dots) are of distance first minimum of lattice $\vec{a}\mZ_q+q\mZ$ to each other. The distance between the two furtherst shifts $\|j\vec{e}\|$ (red dots) has an upper bound, denoted by $d$. Each ball has a radius $R$ chosen to be (approximately) $\lambda_1(\Lambda_q(\vec{a}))/2$, where $\Lambda_q(\vec{a}) = \vec{a}\mZ_q+q\mZ$. Note that once the shaded gray area is measured, the reduction succeeds in outputting an $\GLMSS$ sample. For the reduction to work with a constant success probability, the shaded area has to have a large enough proportion compared to the volume of the balls.
	\label{fig:ball}}
\end{figure}

The same algorithm provides a reduction from $\DLWE$ to $\DGLMSS$. Given a random sample $(\vec{a},\vec{b}) \in \mZ_q^m \times \mZ_q^m$, it suffices to show that all the balls centered at $\vec{a}s+j\vec{b}$ for $s\in \mZ_q$ and $j\in \mZ$, do not intersect with each other. All the points considered above form the lattice $(\vec{a}|\vec{b})\mZ_q+q\mZ$. We argue analogously using the \revdel{upper}\revadd{lower}-bound on the minima of this lattice. As a result, the superposition collapses exactly to one of the balls, which gives a random sample of $\DGLMSS$.

\subsection{Open problems}

\paragraph{Towards an alternative reduction from $\LMSS$ to $\LWE$.}

In~\cite{ChDa07}, Childs and van Dam obtain a state of the form 
\[
\sum_{a \in \Z_N} \sum\limits_{\substack{\vec{j} \in \{0, \ldots, M-1\}^{\ell} \\ \langle\vec{j},\vec{y} \rangle = a \bmod N}} \omega_N^{a \cdot s} \Ket{\vec{j}}.
\] 
for some uniform $\vec{y} \in \Z_N^{\ell}$. Note the uniform distribution of weights for $\vec{j}$. To recover~$s$, 
the authors use the Pretty Good Measurement technique from~\cite{HaWo94} as was 
done in~\cite{BaChDa05,BaChDa06} for similar problems. Implementing this 
general technique to this particular setup requires the construction of
a POVM with operators corresponding to superpositions of \emph{all} the~$\vec{j}$'s
in~$\{0,\ldots,M-1\}^{\ell}$ such 
that~$\langle \vec{j}, \vec{y} \rangle = a \bmod N$. As we already mentioned, a unitary operator that realizes such a POVM, uses a lattice-reduction technique as its main subroutine and, hence, works efficiently only for large values of $M$.

The question we do not address here is the interpretation of the POVM technique (and, possibly, a different reduction to $\LWE$) for Gaussian-weighted superpositions. It might be simpler to obtain Gaussian $\vec{j}$'s rather than uniform from a cube, and hence it is possible that such a technique may lead to an improved reduction to $\LWE$.



\paragraph{Hardness of $\LMSS$ with more input states.}

We show in this work that LWE and $\ULMSS$ are computationally equivalent 
up to small parameter losses, when the number of $\ULMSS$ states~$\ell$ is 
polynomial.
In these reductions, the $\ULMSS$ bound~$M$ is within a polynomial factor 
of the LWE noise rate~$1/\alpha$.
When more states are available, $\ULMSS$ is likely to become easier. For instance,
with~$M=2$, the best known algorithms when~$\ell$ is polynomially bounded are exponential.
Oppositely, Kuperberg's algorithm~\cite{Kuperberg05} runs in  
time~$2^{\widetilde{O}(\sqrt{\log N})}$ 
when~$\ell = 2^{\widetilde{O}(\sqrt{\log N})}$. 
This suggests that there may be a $\ULMSS$ self-reduction
allowing to trade~$\ell$ for~$M$: Is it possible  to reduce 
$\LMSS_{N,\ell,M}$ to $\LMSS_{N,\ell',M'}$ 
with~$\ell' \leq \ell$, while allowing for~$M' \geq M$?

%
%
\def\vd{1.6}
\begin{figure}
	\begin{tikzpicture}
	\coordinate (start) at (-2, 0);
	\node[] (1) at (start) {$\LMSS_{n, q,\sqrt{\kappa}r}^{\ell \kappa^{1.5}}$};
	\node[] (2) at ($(1)+(2.2*\hd, 0)$) {$\LMSS_{n,q, r}^{\ell}$};
	\node[] (3) at ($(2)+(2.2*\hd, 0)$) {$\LMSS_{\mkern3mu n, q, r }^{\mkern3mu \ell/\kappa}$};
	\node[] (3.1) at ($(2)+(2.2*\hd, -0.5)$) {$\LMSS_{\mkern3mu n, q, r }^{\mkern3mu\ell}$};
	\node[] (4) at ($(3)+(2.2*\hd, 0)$) {$\DCP_{q}^{\ell/(\kappa^2 r)}$};
	\node[] (4.1) at ($(3)+(2.30*\hd, -0.5)$) {$\DCP_{q}^{\ell/(\kappa^2 \log r)}$};
	
	\node[] (LWEUp) at  ($(2)+(0, 1.5*\vd)$) {$\LWE_{\substack{\mkern-20mu n, q, \alpha <1/(r m\ell q^{n/m} ) \\ n, q, \alpha <1/(r \sqrt{m}\ell q^{n/m})}}^m$};
	
	\node[] (LWEDown) at  ($(2)+(0, -1.8*\vd)$) {$\LWE_{1, q, \alpha >1/r}^{\ell}$};
	
	\draw[->] (1) -- (2);
	\draw[->] (2) -- (3);
	\draw[->] (3) -- (4);
	
	\draw[->] 
	(1) -- ($(1)+(0,-1.0)$) -- ($(3)+(0,-1.0)$) -> ($(3)+(0,-0.6)$);
	
	\draw[->] 
	($(2)+(0.4, -0.2)$) -- ($(2)+(0.4,-1.5)$) -- ($(4.1)+(-1.0,-1.5)$) -> ($(4.1)+(-1.0,-0.8)$);
	
	\draw[->] (LWEUp)--(2) node[midway, right] {Theorems~\ref{thm:LWEvDCP}, \ref{thm:LWEvDCP_ball}};
	\draw[->] (2)--(LWEDown) node[midway, right, yshift=-1.5em] {Theorem~\ref{thm:vDCPLWE2}};
	

	\node at ($(1)+(1.8,.4)$) {\scriptsize Lemma~\ref{lem:u2g}};
	\node at ($(2)+(1.5,.4)$) {\scriptsize Lemma~\ref{lem:g2u}};	
	\node at ($(3)+(1.6,.4)$) {\scriptsize Lemma~\ref{lem:l2sforlmss}};
	\node at ($(1)+(2.0,-0.8)$) {\scriptsize Lemma~\ref{lem:l2sforlmss}};
	\node at ($(3)+(1.5,-1.3)$) {\scriptsize Lemma~\ref{lem:vDCP2DCP_prime}};

	\end{tikzpicture}
	
	\caption[caption]{Graph of reductions between the extrapolated Dihedral Coset Problem instantiated with uniform distribution over 
	$\{0, 1, \ldots, r-1\}$ (the first and the third problems from the left) and 
	$\ell$-sample Gaussian $\LMSS$ with parameter~$r$ (the middle problem). We assume all the 
	parameters $n$ (the dimension), $q$ (the modulus) and  $r$ are functions of a common 
	parameter~$\kappa$. The most relevant choice of such a relation one can keep 
	in mind is when $n, \ell, q$ and  $r$ are $\poly(\kappa)$.
	One can trace the losses in the parameters (with respect to  the number of 
	samples  $\ell$ and to~$r$) once we move from one problem to another. 
	Notice that some reductions may be performed in two ways. For example, 
	using the self-reducibility property of $\LMSS$ (Lemma~\ref{lem:l2sforlmss}), 
	we can bypass Gaussian $\LMSS$ and have a more sample-efficient reduction 
	from $\LMSS$ with large~$r$ to an $\LMSS$ with smaller~$r$. Similarly, 
	Gaussian $\LMSS$  can be reduced to $\DCP$ either directly 
	(Lemma~\ref{lem:vDCP2DCP_prime}) or via uniform $\LMSS$. \\ 
	The two central reductions that show equivalence between $\LWE$ and $\LMSS$ problems are on the vertical line. As for $\LMSS$, the $\LWE$ parameters $n$, $q$, and $\alpha$ are functions of $\kappa$. We present two reductions from $\LWE$ to $\LMSS$, the stronger one gives a tighter result for the error-parameter by a factor of $\sqrt{m}$.}
	\label{fig:GraphReductionII}
\end{figure}
	

\section{Prerequisites}\label{sec:pre}
\noindent\textbf{Notations.} We use lower case bold letters to denote vectors and upper case bold to denote matrices. For a vector $\vec{x}$, we let $\|\vec{x}\|_\infty$ denote its $\ell_{\infty}$ norm and $\|\vec{x} \|$ denote its $\ell_2$ norm. We let~$\mZ_N$ denote the cyclic group $\{0,1,\cdots,N-1\}$ with addition modulo $N$. We assume we can compute with real numbers. All the arguments are valid if a sufficiently accurate approximation is used instead. 
For a distribution~$D$, the notation $x \hookleftarrow D$ means that $x$ is sampled from $D$. For a set~$S$, we let $x \hookleftarrow S$ denote that $x$ is a uniformly random element from~$S$. 

For any $r>0$, we let $\rho_r(\vec{x})$ denote $\exp(-\pi \|\vec{x}\|^2/r^2)$, where $\vec{x}\in\mR^n$ for a positive integer~$n$. 
We let $\mathcal{D}_{\mZ, r}$ denote a Gaussian distribution over the integers with density function proportional to~$\rho_r(\cdot)$.
We let~$\mathcal{D}_{\Lat,r,\vec{c}}$ denote the Gaussian distribution over the $n$-dimensional lattice~$\Lat$ (for a positive integer~$n$), with standard deviation parameter~$r\in \mR$ and center~$\vec{c}\in \mR^{n}$. If $\vec{c} = \vec{0}$, we omit it. 
We let $\mathcal{B}_n(\vec{c}, R)$ denote the $n$-dimensional Euclidean ball of radius $R$ centered at $\vec{c} \in \R^n$ and $\mathcal{B}_n$ denotes the $n$-dimensional Euclidean unit ball centered at $\vec{0}$. We use $\omega_N$ as a short-hand for $\exp(2 \pi i/N)$.

For a lattice $\Lat$ with a basis $\vec{B}$, the parallelepiped $\mathcal{P}(\vec{B}) = \{\vec{B}\vec{x}: 0\le x_i\le 1\}$ is a fundamental domain of~$\Lat$. We let $\lambda_1(\Lat)$ (resp. $\lambda_1^{\infty}(\Lat)$) denote the $\ell_2$-norm ($\ell_{\infty}$-norm) of a shortest vector of~$\Lat$. We let $\Lat^\star = \{\vec{y} \in \mR^n: \forall \vec{x} \in \Lat, \langle\vec{x},\vec{y}\rangle \in \mZ\}$ denote the dual of a lattice~$\Lat$. 
We define the smoothing parameter $\eta_{\eps}(\Lat)$ as the smallest $r$ such that $\rho_{1/r}(\Lat^\star\backslash \{\vec{0}\}) \le \eps$ for an $n$-dimensional lattice $\Lat$ and $\eps > 0$.

For $\vec{A} \in \Z_q^{m \times n}$, we define two lattices $\qLat (\vec{A}) = \{ \vec{A}\vec{x} \bmod q \colon \vec{x} \in \Z_q^n \}$ and $\qLatp(\vec{A}) = \{\vec{y} \in \Z_q^{n} \; \text{s.t. } \vec{A}\vec{y} = \vec{0} \bmod q\}$. 


 We introduce a variable $\kappa$ to relate all the parameters involved in the definitions below. Namely, $n$, $q$, etc.\ are actually functions in $\kappa$: $n(\kappa)$, $q(\kappa)$. We omit the variable $\kappa$ for clarity. 

\begin{definition}[Search LWE]\label{def:SLWE}
Given a parameter $\kappa$, the input to the search $\LWE_{n,q,\chi}^m$ with dimension $n \ge 1$, modulus $q\ge 2$ and distribution $\chi$ over $\mZ$, consists of $m \ge n$ many samples of the form $(\vec{a}, b) \in \mZ_q^n \times \mZ_q$, with $\vec{a}\hookleftarrow \mZ_q^n$, $b = \langle\vec{a},\vec{s}\rangle + e$ and $e \hookleftarrow \chi$, where $s\in \mZ_{q}^{n}$ is uniformly chosen. We say that an algorithm solves the search $\LWE_{n,q,\chi}^m$ if it outputs $\vec{s}$ with probability $\poly(1/(n\log q))$ in time $\poly(n\log q)$.
\end{definition}

\begin{definition}[Decision LWE]\label{def:DLWE}
Given a parameter $\kappa$, the decisional $\LWE_{n,q,\chi}^m$ with dimension $n \ge 1$, modulus $q\ge 2$ and distribution $\chi$ over $\mZ$, asks to distinguish between $m \ge n$ many $\LWE$ samples and random samples of the form $(\vec{a}, b) \in \mZ_q^n \times \mZ_q$, with $\vec{a}\hookleftarrow \mZ_q^n$, $b \hookleftarrow \mZ_q$. We say that an algorithm solves the decisional $\LWE_{n,q,\chi}^m$ if it succeeds in distinguishing with probability $\poly(1/(n\log q))$ in time $\poly(n\log q)$.
\end{definition}

We let $\LWE_{n,q,\alpha}^m$ (resp.\ $\DLWE_{n,q,\alpha}^m$) denote search (resp.\ decisional) $\LWE$ problem with $m$ samples of dimension $n$, modulus $q$, error distributed as $\mathcal{D}_{\mZ,\alpha q}$.

\begin{definition}[Dihedral Coset Problem]\label{def:DCP}
Given a parameter $\kappa$, the input to the $\DCP_{N}^{\ell}$ with modulus $N$ consists of $\ell$ states. Each state is of the form (normalization is omitted)
\begin{equation}
\Ket{0}\Ket{x} \  + \ \Ket{1}\Ket{(x+s) \bmod N},
\end{equation}
stored on $1+\lceil\log_2 N\rceil$ qubits, where $x \in \mZ_{N}$ is arbitrary and $s\in \mZ_{N}$ is fixed throughout all the states. We say that an algorithm solves $\DCP_{N}^{\ell}$ if it outputs $s$ with probability $\poly(1/\log N)$ in time $\poly(\log N)$.
\end{definition}

Note that Regev in~\cite{Regev02} defines the Dihedral Coset problem slightly differently. Namely, he introduces a failure parameter $f(\kappa)$, and with probability $\leq 1/(\log N(\kappa)^{f(\kappa)})$, we have a state of the form $\Ket{b}\Ket{x}$ for arbitrary $b \in \{0,1\}$ and $x \in \Z_N$. Such a state does not contain any information on $s$. Our definition takes $0$ for the failure parameter. Conversely, Regev's definition is our Def.~\ref{def:DCP} with a reduced number of input states.  

Now we define the problem which can be viewed as an extension of $\DCP$. Analogous to $\LWE$, it has two versions: search and decisional.

\begin{definition}[Search Extrapolated Dihedral Coset Problem]\label{def:LMSS}
Given a parameter $\kappa$, the input to the search Extrapolated Dihedral Coset Problem ($\mkern1mu \LMSS_{n, N, D}^{\ell}$)  with dimension $n$, modulus $N$ and a discrete distribution $D$, consists of $\ell$ input states of the form (normalization is omitted)
\begin{equation}
\sum_{j\in \supp(D)} D(j)\Ket{j}\Ket{(\vec{x}+j\cdot \vec{s}) \bmod N},
\end{equation}
where  $\vec{x} \in \mZ_{N}^n$ is arbitrary and $\vec{s}\in \mZ_{N}^n$ is fixed for all $\ell$ states. We say that an algorithm solves search $\LMSS_{n, N, D}^{\ell}$ if it outputs $\vec{s}$ with probability $\poly(1/(n\log N))$ in time $\poly(n\log N)$.%
%
%
\end{definition}

\begin{definition}[Decisional Extrapolated Dihedral Coset Problem]\label{def:DLMSS}
Given a parameter $\kappa$, the decisional Extrapolated Dihedral Coset Problem $(\mkern-2mu\DLMSS_{\mkern-4mun, N, D}^{\ell}\mkern-2mu)$ with modulus $N$ and a discrete distribution $D$, asks to distinguish between $\ell$~many $\LMSS$ samples and $\ell$~many random samples of the form
\begin{equation}\label{eq:DLMSS_state}
\Ket{j_k}\Ket{\vec{x}_k \bmod N},
\end{equation}
where $j_k \hookleftarrow D^2$ and $\vec{x}_k \in \mZ_{N}^n$ is uniformly chosen for $1 \le k \le \ell$. We say that an algorithm solves $\DLMSS_{n, N, D}^{\ell}$ if it distinguishes the two cases with probability $\poly(1/(n\log N))$ in time $\poly(n\log N)$.
\end{definition}

Different choices of $D$ give rise to different instantiations of $\LMSS$. The two interesting ones are: (1) $D$ is uniform over $\Z_M$ for some $M \in \Z$, which we further denote as $\ULMSS_{n, N, M}^{\ell}$ and (2) $D$ is Gaussian $\gauss_{\mathcal{\Z},r}$, which we further denote as $\GLMSS_{n, N, r}^{\ell}$. The former, named the generalized hidden shift problem, was already considered in \cite{ChDa07}. The latter is central in our reductions. Correspondingly, we call the decisional version of $\GLMSS$ by $\DGLMSS$.

%

\paragraph{Gaussian distribution on lattices.}
In the following, we recall some important properties of discrete Gaussian distribution.

\begin{lemma}\label{lem:gausstail_1}
For any $\kappa, r>0$, we have $\rho_r(\mZ \backslash [-\sqrt{\kappa}r,\sqrt{\kappa}r]) < 2^{-\Omega(\kappa)} \rho_r(\mZ)$.\end{lemma}
For the sake of completeness, we include the proof in Appendix~\ref{pro:gausstail_1}.

From Lemma~\ref{lem:gausstail_1}, we can see that the tail of Gaussian distribution has only negligible proportion compared to the whole sum. We use this fact within a quantum superposition state. For a quantum superposition state with Gaussian amplitudes, the superposition corresponding to Gaussian distribution over full lattice and the one without Gaussian tail have exponentially small $\ell_2$ distance.

\begin{lemma}[{\cite[Lemma 1.5(ii)]{Banaszczyk93}}]\label{lem:shiftgaussiantail}
For any $n$-dimensional lattice $\Lat$ and $\vec{u}\in\mR^n$, it holds that 
\[
\rho_r(\Lat+\vec{u} \backslash \mathcal{B}(\vec{0},\sqrt{n}r)) < 2^{-\Omega(n)} \rho_r(\Lat).
\]
\end{lemma}

\begin{lemma}[Poisson Summation Formula]\label{lem:PSF}
	For any $n$-dimensional lattice $\Lat$ and vector $\vec{u}\in\mR^n$, it holds that
	\[
	\rho_r(\Lambda+\vec{u}) = \mathrm{det}(\Lat^\star)\cdot r^n\cdot \sum_{\vec{x}\in \Lat^\star}e^{2\pi i\langle\vec{x},\vec{u}\rangle}\rho_{1/r}(\vec{x}).
	\]
\end{lemma}

The following Lemma is originally due to Grover-Rudolph \cite{GR02} and was adapted to Gaussian distribution in \cite{Regev09}.

\begin{lemma}[{Adapted from~\cite[Lemma~3.12]{Regev05}}]\label{lem:gausssuperp}
	Given a parameter $\kappa$ and an integer $r$, there exists an efficient quantum algorithm that outputs a state that is within $\ell_2$ distance $2^{-\Omega(\kappa)}$ of the normalized state corresponding to
	\[ 
	\sum_{x\in \mZ}\rho_r(x)\Ket{x}.
	\]
\end{lemma}

The following two lemmata are well-known facts about lower-bounds on minimum of $q$-ary lattices. Proofs of Lemmata~\ref{lem:infbound} and~\ref{lem:l2bound} are given in Appendices~\ref{pro:infbound-v3} and~\ref{pro:l2bound-v3}.

\begin{lemma}\label{lem:infbound}
\sloppy 
Given a uniformly chosen matrix $\vec{A}\in \mZ_q^{m\times n}$ for some positive integers $q$, $m$ and $n$ such that $m \geq n$, we have $\lambda_1^{\infty}(\Lat_q(\vec{A})) \ge q^{(m-n)/m}/4$ with probability at least $1-(1+\ln q)(3/4)^{m}$.
\end{lemma}

\begin{lemma}\label{lem:l2bound}
\sloppy 
Given a uniformly chosen matrix $\vec{A}\in \mZ_q^{m\times n}$ for some positive integers $q$, $m$ and $n$ such that $m \geq 2n$ and $q^{(m-n)/m} \geq 2\sqrt{2\pi e}$, we have $\lambda_1(\Lat_q(\vec{A})) \ge \min \{q, \sqrt{m}\, q^{(m-n)/m}/(2\sqrt{2\pi e})\}$ with probability at least $1-\tau(q)(3/4)^{m} - 3(2m+1)^{15}2^{-m/2}$, where $\tau(q)$ denotes the number of divisors of~$q$.
\end{lemma}
We recall the well-known fact that $\tau(q) = q^{o(1)}$.

\paragraph{Reductions between $\LMSS$ variants.} In the following, we show that the $\LMSS$ problem is analogous to the $\LWE$ problem in many aspects: (1) Gaussian-$\LMSS$ ($\GLMSS_{n,N,r}^{\ell'}$) and uniform-$\LMSS$ ($\ULMSS_{n,N,M}^{\ell}$) are equivalent, up to small parameter losses; (2) $\LMSS$ enjoys the self-reduction property as we show in Lemma~\ref{lem:l2sforlmss}. The main ingredient in both proofs is quantum rejection sampling due to Ozols et al.~\cite{ORR13}.

\begin{lemma}[{\cite[Sec.\ 4]{ORR13}}]\label{lem:qrejsamp}
There is a quantum rejection sampling algorithm, which given as input 
\[
\sum_{k=1}^n \pi_k \Ket{k}\Ket{\eta_k},
\]
for some probability $\pi_k$, outputs 
\[
\frac{1}{\|\vec{p}\|}\sum_{k=1}^n p_k \Ket{k}\Ket{\eta_k}.
\]
for some $p_k \le \pi_k$, with probability $\|\vec{p}\|^2 = \sum_{k=1}^n p_k^2$.
\end{lemma}

\begin{lemma}[$\GLMSS \leq \ULMSS$]\label{lem:g2u}
Let $N, n$ and $\ell$ be integers greater than 1, $r$ be any real number, and let $M = c\cdot r$ for some constant $c$ such that $M$ is an integer. Then there is a probabilistic reduction with run-time polynomial in $\kappa$, from $\GLMSS_{n,N,r}^{\ell}$ to $\ULMSS_{n,N,M}^{\bigO(\ell/\kappa)}$.
\end{lemma}

\begin{proof}
We are given as input $\GLMSS_{n,N,r}^{\ell}$ states:
\[
\left\{\sum_{j\in \mZ}\rho_r(j)\Ket{j}\Ket{(\vec{x}_k+j\cdot \vec{s}) 
\bmod N}\right\}_{k\le \ell}.
\]
Our aim is to find $\vec{s}$, given access to a $\ULMSS_{n,N,cr}^{\bigO(\ell/\kappa)}$ oracle for some constant~$c$.

For each $\GLMSS_{n,N,r}$ sample, we proceed as follows. We let $sign(x)$ denote the sign of $x$, its output is either $1$ (for $\mlq+\mrq$) or $0$ (for $\mlq-\mrq$). We first compute the sign of the first register and store it in a new register:
\[
\sum_{j\in \mZ}\rho_r(j)\Ket{j}\Ket{(\vec{x}+j\cdot \vec{s}) 
\bmod N}\Ket{sign(j)}.
\]

Second, we measure the third register. Note that we observe $1$ with probability at least $1/2$, independently over all $k$'s. If the observed value is $0$, we discard the state. From states with the observed value $1$, we obtain (up to normalization):
\[
\sum_{j\in \Z_+}\rho_r(j)\Ket{j}\Ket{(\vec{x}+j\cdot \vec{s}) 
\bmod N}.
\]

Using quantum rejection sampling  (Lemma~\ref{lem:qrejsamp}), we transform a $\GLMSS_{N,\ell,r}$ state into a $\ULMSS_{n,N,M}$ state of the form
\[
\sum_{j\in [0,M-1]}\Ket{j}\Ket{(\vec{x}+j\cdot \vec{s}) \bmod N}
\]
with probability $\Omega(M\rho_r^2(c\cdot r)/r) = \Omega(1)$.

We repeat the above procedure until we obtain $\bigO(\ell/\kappa)$ many $\ULMSS_{n,N,M}$ states, which happens with probability $\ge 1-2^{-\Omega(\kappa)}$. We call the $\ULMSS_{n,N,M}^{\bigO(\ell/\kappa)}$ oracle to recover the secret $\vec{s}$ as the solution for the input $\GLMSS_{n,N,r}^{\ell}$ instance.
\end{proof}

\begin{lemma}[$\ULMSS \leq \GLMSS$]\label{lem:u2g}
Let $N$, $M$, $n$ and $\ell$ be integers greater than 1, $r$ be any real number, such that $M = \sqrt{\kappa}\cdot r = \poly(\kappa)$ is an integer. Then there is a probabilistic reduction with run-time polynomial in $\kappa$, from $\ULMSS_{n,N,M}^{\ell}$ to $\GLMSS_{n,N,r}^{\bigO(\ell/\kappa^{1.5})}$.
\end{lemma}


\begin{proof}
We are given as input $\ell$ many $\ULMSS_{n,N,M}^{\ell}$ states:
\[
\left\{\sum_{j\in [0,M-1]}\Ket{j}\Ket{(\vec{x}+j\cdot \vec{s}) \bmod N}\right\}_{k\le \ell}.
\]
Our aim is to find $\vec{s}$, given access to a $\GLMSS_{n,N,r}^{\bigO(\ell/\kappa^{1.5})}$ oracle where $r = M/\sqrt{\kappa}$.

For each  $\ULMSS_{n,N,M}$~state we proceed as follows. First, we symmetrize the uniform distribution by applying the function $f(x) = x-\lfloor(M-1)/2\rfloor$ to the first register:
\[
\sum_{j\in [0,M-1]}\mket{j-\lfloor(M-1)/2\rfloor}\Ket{(\vec{x}+j\cdot \vec{s}) \bmod N} = 
\hspace{-1.1cm}
\sum_{j'\in \big[-\lfloor\frac{M-1}{2}\rfloor,\lceil\frac{M-1}{2}\rceil\big]}
\hspace{-1.1cm}
\Ket{j'}\Ket{(\vec{x}'+j'\cdot \vec{s}) \bmod N},
\]
where $j' = j-\lfloor(M-1)/2\rfloor$, $\vec{x}' = \vec{x}+\text{\revdel{$\lceil$}}\text{\revadd{$\lfloor$}}(M-1)/2\text{\revdel{$\rceil$}}\text{\revadd{$\rfloor$}}\cdot \vec{s}$.

Using rejection sampling (Lemma~\ref{lem:qrejsamp}), with probability $\Omega(r/M) = \Omega(1/\sqrt{\kappa})$ we transform each $\ULMSS_{n,N,\lceil\frac{M-1}{2}\rceil}$~state into a $\GLMSS_{n,N,r}$ state:
\[
\sum_{j'\in \big[-\lfloor\frac{M-1}{2}\rfloor,\lceil\frac{M-1}{2}\rceil\big]}\hspace{-1cm}\rho_r(j')\Ket{j'}\Ket{(\vec{x}'+j'\cdot \vec{s}) \bmod N}.
\]

According to Lemma~\ref{lem:gausstail_1}, the latter is within the
$\ell_2$ distance of $2^{-\Omega(\kappa)}$ away from the state
\[
\sum_{j'\in \mZ}\rho_{r}(j')\Ket{j'}\Ket{(\vec{x}'+j'\cdot \vec{s}) \bmod N}.
\]

We repeat the above procedure until we obtain $\bigO(\ell/\kappa^{1.5})$ many $\GLMSS_{n,N,r}$ states, which happens with probability $\ge 1-2^{-\Omega(\kappa)}$. Then we can use the $\GLMSS_{n,N,r}^{\bigO(\ell/\kappa^{1.5})}$ oracle to recover the secret $\vec{s}$ as the solution to $\ULMSS_{n,N,M}^{\ell}$.
\end{proof}

Next, we show the self-reducibility property  for $\LMSS$. We refer the reader to Appendix~\ref{app:l2sforlmss} for the proof.

\begin{lemma}[$\LMSS$ self-reduction]\label{lem:l2sforlmss}
Let $N, n$, and $\ell$ be integers greater than $1$, $r_1$ and $r_2$ be  such that $r_1 > r_2$ and $r_1/r_2 = \bigO(\kappa^c)$ for any constant $c$. Then there is a probabilistic reduction with run-time polynomial in $\kappa$, from $\GLMSS_{n,N,r_1}^{\ell}$ (resp.\ $\ULMSS_{n,N,r_1}^{\ell}$) to $\GLMSS_{n,N,r_2}^{\bigO(\ell/\kappa^{c+1})}$ (resp.\ $\ULMSS_{n,N,r_2}^{\bigO(\ell/\kappa^{c+1})}$).
\end{lemma}

In the following, we give a reduction from Gaussian-$\LMSS$ to $\DCP$. Thus uniform-$\LMSS$ can also be reduced to $\DCP$ in two ways: either via self-reduction, or via Gaussian-$\LMSS$ as the next lemma shows. This result is especially interesting when the parameter $r$ (or $M$ for the uniform-$\LMSS$) is super-polynomially large, as in this case, Lemma~\ref{lem:l2sforlmss} cannot be applied. 
Lemma below works with 1-dimensional $\LMSS$. This is without loss of generality as we can combine our main result (equivalence of $\LWE$ and $\LMSS$) with the result of Brakerski et al.\ \cite{BLPRS13} (equivalence of $\LWE_{n, q, \alpha}$ and $\LWE_{1, q^n, \alpha}$).

\begin{lemma}[Gaussian-$\LMSS$ to $\DCP$]\label{lem:vDCP2DCP_prime}
Let $N$ and $\ell$ be arbitrary integers.
Then there is a probabilistic reduction with run-time polynomial in $\kappa$,  from $\GLMSS_{1,N,r}^{\ell}$ to $\DCP_{N}^{\bigO(\ell/(\log r\cdot \kappa^2))}$ if $r \geq 3\log N$, and from $\GLMSS_{1,N,r}^{\ell}$ to $\DCP_{N}^{\bigO(\ell/(r\cdot \kappa))}$ otherwise.
\end{lemma}

\begin{proof}
We are given as input $\ell$ many $\GLMSS_{1,N,r}$ states:
\[
\left\{\sum_{j\in \mZ}\rho_r(j)\Ket{j}\Ket{(x_k+j\cdot s) 
\bmod N}\right\}_{k\le \ell}.
\]

We show how to find $s$ if we are given access to a $\DCP_{N}^{\bigO(\ell/(r\cdot \kappa))}$ oracle for $r < 3\log N$, and a $\DCP_{N}^{\bigO(\ell/(\log r\cdot \kappa^2))}$ oracle otherwise.

\noindent $\bullet$ Case $r \ge 3\log N$.

According to Lemma~\ref{lem:g2u}, we can transform $\ell$~many $\GLMSS_{1,N,r}$ states into $\ell/\kappa$~many $\ULMSS_{1, N, M'}$ states with $M' = 2 c\cdot r + 1$ for some constant $c$ losing a factor of $\kappa$ samples. Assume we obtain $\ell/\kappa$ many $\ULMSS_{1,N,M'}$ samples. For each such state, we symmetrize the interval $[0, M']$ as in the proof of Lemma~\ref{lem:u2g}. Then we receive a uniform distribution over $[-M, M]$ for $M = (M'-1)/2$. We compute the absolute value of the first register 
and store it in a new register:
\begin{equation}\label{eq:SignState}
\sum_{j\in [-M,M]}\Ket{j}\Ket{(\hat{x}_k+j\cdot s) \bmod N}\Ket{|j|},
\end{equation}
where $\hat{x}_k = x_k - M\cdot s$. We measure the third register and denote the observed value by~$v_k$.

We make use of the two well-known facts from number theory. For proofs, the reader may consult \cite[Chapter~5]{Sho}. First, there exist more than $M/\log M$ many primes that are smaller than $M$. Second, $N$ has at most $2 \log N/ \log\log N$ prime factors. Thus there are at least $M/\log M - 2\log N / \log\log N$ many numbers smaller than $M$ that are co-prime with all prime factors of $N$.

From the above, with probability $\Omega(1/\log M) = \Omega(1/\log r)$, the observed value $v_k$ is non-zero and co-prime with $N$. If this is not the case, we discard the state. Otherwise, we obtain (up to normalization):
\[
\Ket{-v_k}\Ket{(\hat{x}_k-v_k\cdot s) \bmod N} 
+ \Ket{v_k}\Ket{(\hat{x}_k+v_k\cdot s) \bmod N}.
\]

We multiply the value in the second register by $v_k^{-1} \bmod N$:
\[
\Ket{-v_k}\Ket{(x_k'-s) \bmod N} 
+ \Ket{v_k}\Ket{(x_k'+s) \bmod N},
\]
where $x_k' = \hat{x}_k\cdot v_k^{-1}$.

Let $\bar{x}_k = x_k'-s \bmod N$ and $\bar{s} = 2 \cdot s \bmod N$. Rewrite the above state as:
\[
\Ket{-v_k}\Ket{\bar{x}_k} + \Ket{v_k}\Ket{(\bar{x}_k+\bar{s}) \bmod N}.
\]

As we know $v_k$ classically, we uncompute the first register and obtain a $\DCP$ state:
\begin{equation}\label{eq:DCPState}
\Ket{0}\Ket{\bar{x}_k} + \Ket{1}\Ket{(\bar{x}_k+\bar{s}) \bmod N}.
\end{equation}


We repeat the above procedure until we obtain $\bigO(\ell/(\log r \cdot \kappa^2))$ many $\DCP_{N}$ states with probability $\ge 1-2^{-\Omega(\kappa)}$. 

\noindent $\bullet$ Case that $r < 3\log N$.

The first steps are identical to the proof for the case $r\geq 3 \log N$: Compute the absolute value of the first register to get a state as in~\eqref{eq:SignState} and measure the third register. Denote the observed value by~$v_k$. Now we keep only those states, for which $v_k=1$ was observed. Otherwise, we do not use the state. In case $v_k=1$, we can easily transform the result to the state given in~\eqref{eq:DCPState} analogously to the proof for $r \geq 3 \log N$.

Now we show that $v_k= 1$ occurs with probability $\Omega(1/r)$ 
independently over all $k$'s. Indeed,
\[
\Pr[v_k=1] = \frac{\rho_r(1)^2 + \rho_r(-1)^2}{\sum_{j\in\mZ}\rho_r(j)^2} \geq \frac{2\cdot \rho_r(1)^2}{\int_{\mR}\rho_r(x)^2 \mathrm{d}x + 1} = \frac{2\cdot \exp(-\frac{2\pi}{r^2})}{\frac{r}{\sqrt{2}}+1} = \Omega\Big(\frac{1}{r}\Big).
\]
%
We  repeat the above procedure until we obtain $\bigO(\ell/(r\cdot \kappa))$ many
$\DCP_{N}$ states, which happens with probability~$\geq 1-2^{-\Omega(\kappa)}$.

Therefore, if $r\geq 3\log N$, we use the  $\DCP_{N}^{\bigO(\ell/(\log r\cdot \kappa^2))}$ oracle, and the $\DCP_{N}^{\bigO(\ell/(r\cdot \kappa))}$ oracle otherwise, to get the secret~$\bar{s}$. There are at most $2$ possible values $s$ such that $\bar{s} = 2s\bmod N$: if there are $2$ possibilities, we uniformly choose either, which decreases the success probability by at most a factor of~$2$.
\end{proof}


\section{Reduction from $\LWE$ to $\LMSS$}\label{sec:LWEvDCP}
In this section, we reduce $\LWE_{n,q,\alpha}^{m}$ to 
$\GLMSS_{n,q,r}^{\ell}$, where $r \approx 1/\alpha$ up to a factor of $\mathrm{poly}(n \log q)$. 
Analogous to Regev's reductions from $\uSVP$ to $\DCP$, we present two versions of the reduction from $\LWE$ to 
$\GLMSS$. The second one is tighter with respect to the parameter losses. 
At the end of the section we show that using the same algorithm, one can reduce the decisional version of $\LWE$ to the decisional version of $\LMSS$ (see Def.~\ref{def:DLMSS}). 

\subsection{First reduction: using cube separation}
The main result of this section is the following theorem.
\begin{theorem}[$\LWE \leq \LMSS$]\label{thm:LWEvDCP}
Let $(n, q, \alpha)$ be $\LWE$ parameters with $q \ge 3$ and $(n, q, r)$ be $\LMSS$ parameters. Given $m = n\log q = \Omega(\kappa)$~many $\mathrm{LWE}_{n,q,\alpha}$ samples, there exists a probabilistic quantum reduction, with run-time polynomial in $\kappa$, from $\mathrm{LWE}_{n,q,\alpha}^{m}$ to $\GLMSS_{n,q,r}^{\ell}$, where $r < 1/(128m\kappa\alpha\ell q^{n/m})$.
\end{theorem}

The main step of our reduction is to partition the ambient space $\R^m$ with an appropriately chosen grid (cubes). This is analogous to Regev's reduction from $\uSVP$ to $\DCP$~\cite{Regev02}. Lemma~\ref{lem:seperate_gaussian_ball} shows how we choose the width of the cell in our grid. Figure~\ref{fig:grid} gives a 2-dimensional example of such a grid. 
\begin{lemma}\label{lem:seperate_gaussian_ball}
Let $c \ge 8$, $q$, $n$, $m = n\log q$ and $k \ge m$ be integers, and let $\vec{A} \in \mZ_q^{m\times n}$ be such that the map $\vec{x} \mapsto \vec{A}\vec{x} \bmod q$ is injective on~$\mZ_q^n$. We
consider a function
\[
g: (x_1, \cdots, x_m) \rightarrow (\lfloor x_1/z +w_1 \bmod \bar{q}\rfloor, \cdots, \lfloor x_m/z+w_m \bmod \bar{q}\rfloor),
\]
where $z = q/c$ and $z \in [1/c,1/2]\cdot \lambda_1^{\infty}(\Lambda_q(\vec{A}))$, $w_1,\ldots,w_m$ are uniformly chosen from $[0,1)$, and $\bar{q}= q/z$. Then for any $\vec{x}\in \mZ_q^{n}$, we have the following two statements.
\begin{itemize}
  \item For any $\vec{u} = \vec{A} \vec{x} + \vec{e}_1, \vec{v} = \vec{A} \vec{x} + \vec{e}_2$ where $\|\vec{e}_1\|_{\infty}, \|\vec{e}_2\|_{\infty} \le \lambda_1^{\infty}(\Lambda_q(\vec{A}))/(2ck)$, with probability at least $(1-1/k)^m$, over the randomness of $w_1,\cdots,w_m$, we have $g(\vec{u}) = g(\vec{v})$.
  \item  For any $\vec{u} = \vec{A} \vec{x} + \vec{e}_1, \vec{v} = \vec{A} \widehat{\vec{x}} + \vec{e}_2$, where $\|\vec{e}_1\|_{\infty}, \|\vec{e}_2\|_{\infty} \le \lambda_1^{\infty}(\Lambda_q(\vec{A}))/(2ck)$ and $\vec{x} \ne \widehat{\vec{x}} \in \mZ_q^{n}$, we have $g(\vec{u}) \ne g(\vec{v})$.
\end{itemize}
\end{lemma}

\begin{proof}

$\bullet$ Proof for the first claim.

Write $\vec{u} = \vec{A} \vec{x}+\vec{e}_1 \bmod q$ and $\vec{v} = \vec{A} \vec{x}+\vec{e}_2 \bmod q$ for some $\vec{x}\in \mathbb{Z}_q^{n}$ and $\|\vec{e}_1\|_{\infty}, \|\vec{e}_2\|_{\infty} \le \lambda_1^{\infty}(\Lambda_q(\vec{A}))/(2ck)$. 

Let $\textsc{diff}$ denote the event that $g(\vec{u}) \ne g(\vec{v})$, 
and, for all $i \leq m$, let $\textsc{diff}_i$ denote the event that 
the $i\th$ coordinates of $g(\vec{u})$ and $g(\vec{v})$ differ. Since we choose $w_1, \ldots, w_m$ independently and uniformly from $[0, 1)$, we can consider each of $m$ dimensions separately and view each $e_{1, i}/z+w_i$ and $e_{2, i}/z+w_i$ as random $1$-dim.\ real points inside an interval of length 1. We have
\[
\Pr_{w_i}[\textsc{diff}_i] = \frac{|e_{1,i}-e_{2,i}|}{z} \le \frac{z/k}{z} = 
\frac{1}{k},
\]
where the inequality follows from the lower-bound on $z$. This implies
\[
\Pr_{\vec{w}}[\textsc{no diff}] = \prod_{i\leq m}\left(1-\Pr_{w_i}[\textsc{diff}_i]\right) \ge \left(1-\frac{1}{k}\right)^m.
\]

$\bullet$ Proof for the second claim.

Write $\vec{u} = \vec{A} \vec{x}+\vec{e}_1\bmod q$ and $\vec{v} = \vec{A} \widehat{\vec{x}}+\vec{e}_2\bmod q$ for $\vec{x} \ne \widehat{\vec{x}} \in \mZ_q^{n}$ and 
$\|\vec{e}_1\|_{\infty}, \|\vec{e}_2\|_{\infty} \le \lambda_1^{\infty}(\Lambda_q(\vec{A}))/(2ck)$. Then we have
\begin{eqnarray*}
g(\vec{u})
 &=& \Big\lfloor \frac{1}{z}\cdot(\vec{A} \vec{x}) + \frac{1}{z}\cdot\vec{e}_1 + \vec{w} \bmod \bar{q}\Big\rfloor, \\
g(\vec{v})
&=& \Big\lfloor \frac{1}{z}\cdot(\vec{A} \widehat{\vec{x}}) + \frac{1}{z}\cdot\vec{e}_2 + \vec{w} \bmod \bar{q}\Big\rfloor.
\end{eqnarray*}


Now we show that $g(\vec{u})$ and $g(\vec{v})$ differ in at least 1 coordinate. This is the case if the arguments of the floor function differ by 1 in at least one coordinate, i.e., $\| \tfrac{1}{z} \vec{A}\cdot (\vec{x}-\hat{\vec{x}}) + \tfrac{1}{z}  (\vec{e}_1 - \vec{e}_2) \bmod \bar{q} \|_{\infty} \geq 1$. 

Assume the contrary is the case.  Note that due to our choice of $\vec{e}_i$ and $\bar{q}$, $\| \tfrac{1}{z}  (\vec{e}_1 - \vec{e}_2) \bmod \bar{q} \|_{\infty}$ is either at most $1/k$ or at least  $\bar{q}-1/k$. Either way we have $\| \tfrac{1}{z} \vec{A} (\vec{x}-\hat{\vec{x}}) \bmod \bar{q} \|_{\infty} < 1 + 1/k$
or $\|\tfrac{1}{z} \vec{A} (\vec{x}-\hat{\vec{x}}) \bmod \bar{q} \|_{\infty} > \bar{q}  - 1- 1/k$. 
Due to the bounds on $z$ and $c$, the former case is equivalent to
\[
	\|\vec{A}(\vec{x}-\hat{\vec{x}}) \bmod q \|_{\infty} < z + z/k \leq \lambda_1^{\infty}(\Lambda_{q}(\vec{A})) \Big(\tfrac{1}{2} + \tfrac{1}{2k} \Big) \leq \lambda_1^{\infty}(\Lambda_{q}(\vec{A})).
\]
As $\vec{x} \ne \hat{\vec{x}}$ and $\vec{x} \mapsto \vec{A}\vec{x} \bmod q$ is injective, we have $\vec{A}(\vec{x}-\hat{\vec{x}}) \not\equiv \vec{0} \bmod q$. Hence, we have just found a nonzero vector in the lattice $\Lambda_{q}(\vec{A})$ shorter than the minimum of the lattice. In the latter case when $\| \tfrac{1}{z} \vec{A}\cdot (\vec{x}-\hat{\vec{x}}) \bmod \bar{q} \|_{\infty} >\bar{q}-1-1/k$, we obtain the same contradiction by noticing that $\Lambda_q(\vec{A})$ contains $q$-ary vectors.
\end{proof}
\begin{figure}
\[
\Qcircuit @C=1.0em @R=.5em @!R {
	\lstick{\ket{0}} & \qw & \qw & \qw \ar@{--}+<0.1em, 1em>;[4,0]+<0.1em,-0.8em>&  \gate{\parbox{1cm}{\centering \tiny Grover\\ Rudolph}} & \qw \ar@{--}+<0.1em, 1em>;[4,0]+<0.1em,-0.8em>  & \multigate{2}{U_f} \ar@{--}+<1.6em, 1em>;[4,0]+<1.6em,-0.8em> & \qw \ar@{--}+<1.4em, 1em>;[4,0]+<1.4em,-0.8em>  & \qw & \qw & \multigate{2}{U_f} & \qw & \qw \ar@{--}+<0.7em, 1em>;[4,0]+<0.7em,-0.8em> & \multigate{1}{U_{\LMSS}} \\
	\lstick{\ket{0}} & \qw & \gate{QFT_{\Z_q^n}} & \qw & \qw & \qw & \ghost{U_f} & \qw & \qw & \qw & \ghost{U_f} & \qw & \qw  &\ghost{U_{\LMSS}} \\
	\lstick{\ket{0}} & \qw & \qw & \qw & \qw & \qw & \ghost{U_f} & \multigate{1}{U_g} & \qw & \qw & \ghost{U_f} & \rstick{\ket{0}} \qw & &   \\
	\lstick{\ket{0}} & \qw & \qw & \qw & \qw & \qw & \qw & \ghost{U_g} & \qw & \meter  &  \measureD{\scalebox{0.6}{$g(\vec{a}\cdot s-j\vec{e}_0)$}} & \\
	& & & & & & & & & & & &\\
	& & &\eqref{eq:InpStateThm2} & & \eqref{eq:GaussStateThm2} &  \hspace{3em}\eqref{eq:StateU_fThm2} &\hspace{3.0em} \eqref{eq:StateU_gThm2} & & & & &\hspace{1.0em}\eqref{eq:ULMSSThm2} \\
}
\]
\caption{Quantum circuit for our reduction $\LWE$ $\leq$ $\LMSS$. All the global phases are omitted. The input registers are assumed to have the required number of qubits. Function $f$ is defined as $U_f\ket{j}\ket{\vec{s}}\ket{0} \rightarrow \ket{j}\ket{\vec{s}}\ket{\vec{A} \vec{s}-j\vec{b} \bmod q}$. Function $U_g$ is the embedding of function $g$ described in Lemma~\ref{lem:seperate_gaussian_ball}, i.e.\ $U_g\ket{\vec{x}}\ket{0} \rightarrow\ket{\vec{x}}\ket{\lfloor \vec{x}/z +\vec{w} \bmod \bar{q} \rfloor}$ for appropriately chosen $z, \vec{w}, \bar{q}$. }
\label{fig:LWEtoLMSS}
\end{figure}

\begin{proof}[of Theorem~\ref{thm:LWEvDCP}]
Assume we are given an $\LWE_{n,q,\alpha}^{m}$ instance
$(\vec{A},\vec{b}_0)$ with 
$\vec{b}_0=\vec{A}\cdot \vec{s}_0+\vec{e}_0 \bmod q$. 
Our aim is to find $\vec{s}_0$ given access to a $\GLMSS_{n,q,r}^{\ell}$ oracle.

We first prepare a necessary number of registers in the state $\Ket{0}$ and transform them to the state of the form (normalization omitted)
\begin{equation} \label{eq:InpStateThm2}
\sum_{\vec{s} \in \mZ_q^{n}} \Ket{0} \Ket{\vec{s}} \Ket{0}.
\end{equation}

We use Lemma~\ref{lem:gausssuperp}  to obtain a state 
within $\ell_2$ distance of $2^{-\Omega(\kappa)}$ away from 
\begin{equation}\label{eq:GaussStateThm2}
\sum_{\vec{s} \in \mZ_q^{n}} \mkern-7mu\Big(\sum_{j \in \mZ}\rho_{r}(j)\Ket{j}\Big)\Ket{\vec{s}}\Ket{0}.
\end{equation}

According to Lemma~\ref{lem:gausstail_1}, the state above is within 
$\ell_2$ distance of $2^{-\Omega(\kappa)}$ away from\footnote{Here we cut the tail of the Gaussian distribution on the first register. Otherwise, a measurement that follows leads to a state mixed with noisy vectors from different lattice points with large (unbounded) noise. However, it has $\ell_2$ distance exponentially close to the state we consider in the current algorithm.}
\[
\sum_{\substack{\vec{s} \in \mZ_q^{n} \\ j \in \mZ \cap [-\sqrt{\kappa} \cdot r,\sqrt{\kappa}\cdot r]}} \hspace*{-.9cm}
\rho_{r}(j)\Ket{j}
\Ket{\vec{s}}\Ket{0}.
\]

We evaluate the function $f(j,\vec{s}) \mapsto  \vec{A} \vec{s} - j\cdot \vec{b} \bmod q$ and store the result in the third register. The next equality follows from a change of variable on $s$
\begin{equation*} \label{eq:StateU_fThm2}
    \sum_{\substack{ \vec{s} \in \mZ_q^{n} \\ j \in \mZ \cap [-\sqrt{\kappa} \cdot r,\sqrt{\kappa}\cdot r]}} \hspace*{-.8cm} 
		\rho_{r}(j)\Ket{j}\Ket{\vec{s}}\Ket{\vec{A} \vec{s}-j \cdot \vec{A} \vec{s}_0-j \vec{e}_0} \ = \hspace*{-.9cm}
		\sum_{\substack{ \vec{s} \in \mZ_q^{n} \\ j \in \mZ \cap [-\sqrt{\kappa} \cdot r,\sqrt{\kappa}\cdot r]}} \hspace*{-.8cm} 
		\rho_{r}(j)\Ket{j}\Ket{\vec{s}+j \vec{s}_0} \Ket{\vec{A} \vec{s}-j \vec{e}_0}.
\end{equation*}

Sample $w_1,\ldots,w_m$  uniformly from~$[0,1)$.
Set $z = q/c$ with $c = 16$, thus we have $z \in [1/c,1/2]\cdot \lambda_1^{\infty}(\Lambda_q(\vec{A}))$, where the upper bound holds with probability at least $1-(1+\ln q)(3/4)^m$ $= 1-2^{-\Omega(\kappa)}$ (see Lemma~\ref{lem:infbound} and note that $q^{(m-n)/m} = q/2$). Further, the map $\vec{x} \mapsto \vec{A}\vec{x} \bmod q$ is injective on~$\mZ_q^n$, as required in Lemma~\ref{lem:seperate_gaussian_ball}, with probability at least 
\[
1-\sum_{d \mid q,\, d>1} d^{n-m} \ge 1-3\cdot 2^{n-m} = 1-2^{-\Omega(\kappa)},
\]
as $m-n \ge m/3$ for $q \ge 3$. Indeed, for each divisor $d>1$ of~$q$, there are at most $d^n$ vectors $\vec{x}$ with $\gcd(x_1,\ldots,x_n,q) = q/d$, and each of them satisfies $\vec{A}\vec{x} \equiv \vec{0} \bmod q$ with probability~$d^{-m}$. Moreover, comparing with an integral and using $m-n \ge 2$, we have $\sum_{d \mid q,\, d>1} d^{n-m} \le \sum_{k \ge 2} k^{n-m} \le 2^{n-m} + \int_2^{\infty} x^{n-m}\,\mathrm{d}x = \big(1+\frac{2}{m-n-1}\big)\, 2^{n-m} \le 3\cdot 2^{n-m}$.

For $\vec{x} \in \mZ_q^m$, we define
\[
g(\vec{x}) = (\lfloor (x_1/z +w_1) \bmod \bar{q}\rfloor,\ldots,\lfloor (x_m/z+w_m) \bmod \bar{q}\rfloor),
\]
where $\bar{q}=q/z=c$. We evaluate the function $g$ on the third register and store the result on a new register. We obtain
\begin{equation} \label{eq:StateU_gThm2}
\sum_{\substack{ \vec{s} \in \mZ_q^n \\ j \in \mZ \cap [-\sqrt{\kappa} \cdot r,\sqrt{\kappa}\cdot r]}} \hspace*{-.8cm} 
\rho_{r}(j)\Ket{j}\Ket{\vec{s}+j\cdot \vec{s}_0}\Ket{\vec{A} \vec{s}-j\cdot \vec{e}_0}\Ket{g(\vec{A} \vec{s}-j\cdot \vec{e}_0)}.
\end{equation}

We measure the fourth register and do not consider it further. 
According to Lemma~\ref{lem:gausstail_1}, we have $\|\vec{e}_0\|_\infty \le \sqrt{\kappa}\alpha q$ with probability $\ge 1-2^{-\Omega(m)} = 1-2^{-\Omega(\kappa)}$. Recall that $r < 1/(128 m\ell\kappa\alpha q^{n/m}) = 1/(8ck\kappa\alpha q^{n/m})$ for $c = 16$ and $k = m\ell$. As $\lambda_1^{\infty}(\Lambda_q(\vec{A})) \ge q^{(m-n)/m}/4$, we have $\|\sqrt{\kappa}r\cdot \vec{e}_0\|_\infty \le \lambda_1^{\infty}(\Lambda_q(\vec{A}))/(2ck)$. Then by Lemma~\ref{lem:seperate_gaussian_ball} applied to the two extreme shifts $j = \pm\lfloor\sqrt{\kappa}r\rfloor$ (the intermediate shifts lie coordinatewise between them, hence in the same cells),
we obtain
\[
\sum_{j \in \mZ \cap [-\sqrt{\kappa} \cdot r,\sqrt{\kappa}\cdot r]} 
\hspace*{-.3cm}
\rho_r(j)\Ket{j}\Ket{\vec{s}+j\cdot \vec{s}_0}\Ket{\vec{A} \vec{s}-j\cdot \vec{e}_0} 
\] 
for some $\vec{s} \in \mZ_q^{n}$, with probability at least $(1-1/k)^m$ over the randomness of $\vec{A}$ and $w_1, \cdots, w_m$.

Finally, we evaluate the function $(j,\vec{s},\vec{b}) \mapsto \vec{b}-\vec{A} \vec{s}+j\cdot\vec{b}_0$ on the first three registers, which gives $\vec{0}$. Discarding this $\vec{0}$-register, the state is of the form
\[
\sum_{j \in \mZ \cap [-\sqrt{\kappa} \cdot r,\sqrt{\kappa}\cdot r]}\hspace*{-.3cm}
\rho_{r}(j)\Ket{j}\Ket{\vec{s}+j\cdot \vec{s}_0}.
\]
According to Lemma~\ref{lem:gausstail_1}, the above state is within 
$\ell_2$ distance of $2^{-\Omega(\kappa)}$ away from
\begin{equation*} \label{eq:ULMSSThm2}
\sum_{j\in \mZ}\rho_{r}(j)\Ket{j}\Ket{\vec{s}+j\cdot \vec{s}_0}.
\end{equation*}

We repeat the above procedure $\ell$ times, and with probability $(1-\frac{1}{k})^{m \ell}$, we obtain $\ell$ many $\GLMSS_{n,q,r}^{\ell}$ states
\[
\left\{\sum_{j\in \mZ}\rho_{r}(j)\Ket{j}\Ket{\vec{x}_k+j\cdot \vec{s}_0}\right\}_{k\le \ell},
\]
where $\vec{x}_k \in \mZ_q^n$.

Now we can call the $\GLMSS_{n,q,r}^{\ell}$ oracle with the above states as input and obtain $\vec{s}_0$ as output of the oracle.
\end{proof}

\subsection{An improved reduction: using balls' intersection}\label{sec:LWEvDCP_ball}
Here we give an improved reduction from $\LWE$ to $\LMSS$. Following the idea of Regev (\cite[Section~3.3]{Regev02}), instead of separating the ambient space $\Z^m$ by cubes, we consider intersections of balls drawn around the points $\vec{A}\vec{s}$ and its shifts. Note that with this reduction we improve the upper-bound on $r$ essentially by the factor of $\sqrt{m}$. 

\begin{theorem}[$\LWE \leq \LMSS$]\label{thm:LWEvDCP_ball}
Let $(n, q, \alpha)$ 
be  $\LWE$ parameters and $(n, q, r)$  be  $\LMSS$ parameters. Given $m = \Omega(\kappa+\log q)$~many $\mathrm{LWE}_{n,q,\alpha}$ samples, where $m \ge 2n$ and $2\sqrt{2\pi e} \le q^{(m-n)/m} \le 3\sqrt{2\pi e}\, q/\sqrt{m}$, there exists a quantum reduction, with run-time polynomial in $\kappa$, from $\mathrm{LWE}_{n,q,\alpha}^m$ to $\GLMSS_{n,q,r}^{\ell}$, where $r < 1/(6\sqrt{2\pi e}\sqrt{m \kappa}\ell\alpha q^{n/m})$.
\end{theorem}

We give an intuitive idea of how the reduction works. All the necessary lemmata and the full proof are given in Appendix~\ref{app:LWEvDCP}.

Informally, the reduction works as follows. Given an $\LWE$ instance $(\vec{A}, \vec{b} = \vec{A}\vec{s}_0+\vec{e}_0) \in \Z_q^{m \times n} \times \Z_q^m$, for each $\vec{s} \in \Z_q^n$, we consider (in a superposition over all such $\vec{s}$) a lattice point $\vec{A}\vec{s}$ together with its small shifts of $\vec{A}\vec{s} - j \vec{e}_0$, where $j$'s are drawn from a small interval symmetric around 0. So far this is exactly what we did in the first (weaker) reduction. Note that we receive a configuration of points in $\Z_q^m$ as depicted in Figure~\ref{fig:ball}. Note that contrary to Regev's reduction, where there is only one shift (i.e., the DCP case), our extrapolated version considers $\poly(\kappa)$ shifts thus leading us to the $\LMSS$ case.

Let us fix some $\vec{A}\vec{s}$ together with its shifts. Draw a ball around each shift of a maximal radius $R$ such that there is no intersection between the  shifts coming from different lattice points, i.e. there is no $j, j'$ s.t.\ $\mathcal{B}(\vec{A}\vec{s} - j \vec{e}_0, R) \cap \mathcal{B}(\vec{A}\vec{s'} - j' \vec{e}_0, R) \neq \emptyset $ for any two $\vec{s}, \vec{s'}$ such that $\vec{s}\ne \vec{s}'$. To satisfy this condition, we can take $R$ almost as large as the first minimum of the lattice $\qLat(\vec{A})$ (again, see Figure~\ref{fig:ball}). With such an $R$, due to the fact that the shifts are small, the intersection of the balls drawn around the shifts is large enough (see Lemma~\ref{lem:sphereintersect} in Appendix~\ref{app:LWEvDCP}). Hence, once we measure the register that `stores' our balls, the resulting state collapses (with large enough probability) to a superposition of some $\vec{A}\vec{s}$ for \emph{one} $\vec{s}$ and all its shifts. Informally, the higher this probability is, the tighter the parameters achieved by the reduction.

\subsection{Reduction from $\DLWE$ to $\DLMSS$} \label{dLWEdLMSS}
As a corollary to the above theorem, we show that the decisional $\LWE$ can be reduced to decisional $\LMSS$. In fact, to establish the reduction, we use the same algorithm as for Theorem~\ref{thm:LWEvDCP_ball} (a weaker reduction given in Theorem~\ref{thm:LWEvDCP} will work as well). Recall that in the proof, starting from an $\LMSS$ sample, we obtain an $\LWE$ sample with non-negligible probability. Corollary~\ref{cor:dLWEvDCP} below shows that in case we are given a tuple $(\vec{A}, \vec{b})$ drawn uniformly at random from $\Z_q^{m \times n} \times \Z_q^m$, the procedure described in Theorem~\ref{thm:LWEvDCP_ball} outputs a state of the form $\Ket{j}\Ket{\vec{x} \bmod N}$, a uniform counterpart to $\LMSS$ in the sense of Definition~\ref{def:DLMSS}. The proof of the following corollary is given in Appendix~\ref{app:LWEvDCP}.

\begin{corollary}[$\DLWE \leq \DLMSS$]\label{cor:dLWEvDCP}
\sloppy 
	Let $(n, q, \alpha)$ 
	be valid  $\DLWE$ parameters and $(n, q, r)$  be valid  $\DLMSS$ parameters.
	 Given $m = \Omega(\kappa+\log q)$~many $\mathrm{LWE}_{n,q,\alpha}$ samples, where $m \ge 2(n+1)$ and $2\sqrt{2\pi e} \le q^{(m-n-1)/m} \le 3\sqrt{2\pi e}\, q/\sqrt{m}$, there exists a quantum reduction, with run-time polynomial in $\kappa$ and~$1/\varepsilon$, from $\DLWE_{n,q,\alpha}^m$ to $\DGLMSS_{n,q,r}^{\ell}$, where $r < \varepsilon/(C\sqrt{m\kappa}\ell\alpha q^{(n+1)/m})$ for some absolute constant $C>0$. Here $\varepsilon$ denotes the distinguishing advantage of the $\DGLMSS$ oracle, and the reduction distinguishes with advantage at least $\varepsilon/2 - 2^{-\Omega(\kappa)}$.
\end{corollary}


\section{Reduction from $\LMSS$ to $\LWE$}\label{sec:vDCPLWE2}
In this section, we reduce $\GLMSS_{n,N,r}^{\ell}$ to 
$\LWE_{n,N,\alpha}^{\ell}$, where $r \approx 1/\alpha$ up to a factor of $\mathrm{poly}(n\log N)$. Combined with the result of the previous section, this gives us equivalence between the two problems: $\LWE$ and $\LMSS$, for both search and decisional variants. 
\begin{figure}[h]
\[
	\hspace{-50pt}
	\Qcircuit @C=1.em @R=.7em @!R {
			& \ar@{--}+<0.2em, 1em>;[2,0]+<0.2em,.5em> \qw & \qw & \qw & \qw \ar@{--}+<0.1em, 1em>;[2,0]+<0.1em,-0.3em> & \qw & \qw  & \qw \ar@{--}+<0.2em, 1em>;[2,0]+<0.2em,-0.3em>&   \gate{\parbox{1.0cm}{\scriptsize $QFT_{\Z_N}$}} & \qw & \hspace*{3.2cm} \sum\limits_{e\in \mZ_N}\rho_{\frac{1}{r}}\Big(\frac{e}{N}\Big) \Ket{\langle\vec{a}_k', \vec{s}_0\rangle + e}\\
			& \qw & \qw  & \gate{\parbox{1.0cm}{\scriptsize $QFT_{\Z_N^n}$}} & \qw & \meter & \measureD{\vec{a}_k} & & & & \\
			& \scalebox{0.8}{$\sum\limits_{j\in \mZ} \rho_r(j) \Ket{j}\Ket{\vec{x}_k+j\cdot \vec{s}_0}$} & & & & & & & &\\
			& & & & \scalebox{0.9}{{\eqref{eq:QFTStep1}}} & & & \scalebox{0.9}{{\eqref{eq:gaussiancut}}} & &\\
		}
\]
	\caption{Reduction from $\GLMSS$ to $\LWE$ } 
	\label{fig:GLMSStoLWE}
\end{figure}

\begin{theorem}[$\LMSS \leq \LWE$]\label{thm:vDCPLWE2}
Let $(n, N, r)$ be valid $\LMSS$ parameters and $(n, N, \alpha)$ with $r = \Omega(\sqrt{\kappa})$
be valid $\LWE$ parameters, with $N > 2\sqrt{\kappa}\, r$. Given $\ell = \Omega(\kappa)$~many $\GLMSS_{n,N,r}$ samples, there exists a quantum reduction, with run-time polynomial in $\kappa$, from $\GLMSS_{n,N,r}^{\ell}$ to $\mathrm{LWE}_{n,N,\alpha}^{\ell}$, where $\alpha = 1/r$.
\end{theorem}

\begin{proof}
Assume we are given $\ell$~many $\LMSS_{n,N,r}$ instances
\[
\left\{\sum_{j\in \mZ} \rho_r(j) \Ket{j}\Ket{\vec{x}_k+j\cdot \vec{s}_0 \bmod N} \right\}_{k\in [\ell]}.
\]
Our aim is to find $\vec{s}_0$ given access to an $\LWE_{n,N,\alpha}^{\ell}$ oracle.

        For each input state, the quantum Fourier transform over $\mZ_N^n$ is applied to the second register, which yields (without loss of generality, consider the $k\th$ sample)
	\begin{equation}
	\label{eq:QFTStep1}
	\sum_{\vec{a}\in \mZ_N^n} 
	\sum_{j\in \mZ} \omega_N^{\langle \vec{a},(\vec{x}_k+j\cdot \vec{s}_0)\rangle}\cdot 
	\rho_{r}(j)\Ket{j} \Ket{\vec{a}}.
	\end{equation}

        Then we measure the second register and let $\vec{a}_k$ denote the observed value. Note that each element of $\mZ_N^n$ is measured with 
	probability~$1/N^n$ and that the distributions for different $k$'s are independent. 
	Omitting the global phase of each state, we obtain
	\begin{equation}
	\label{eq:measure1}
	\sum_{j\in \mZ} \omega_N^{\langle \vec{a}_k, (j\cdot \vec{s}_0) \rangle}\cdot \rho_{r}(j) \Ket{j}
	\Ket{\vec{a}_k}.
	\end{equation}
        We omit the second register as we know each $\vec{a}_k$ classically. 
    Since $N > 2\sqrt{\kappa}\, r$, from Lemma~\ref{lem:gausstail_1} it follows that the resulting state is within 
$\ell_2$ distance of $2^{-\Omega(\kappa)}$ away from the state (note the change in the range for~$j$)
        \begin{equation}
	\label{eq:gaussiancut}
	\sum_{j \in \mZ_N} \omega_N^{j\cdot \langle \vec{a}_k,\vec{s}_0\rangle}\cdot \rho_{r}(j) \Ket{j}.
	\end{equation}
        For each such input state, the quantum Fourier transform over $\mZ_N$ yields
        \begin{equation}
	\label{eq:QFTstep2}
        \sum_{b\in \mZ_N}\sum_{j\in \mZ_N} \omega_N^{j\cdot( \langle \vec{a}_k, \vec{s}_0\rangle + b)}\cdot \rho_{r}(j) \Ket{b}.
        \end{equation}
        Once again we use Lemma~\ref{lem:gausstail_1} to argue that the state above is within $\ell_2$ distance of $2^{-\Omega(\kappa)}$ away from the state
        \begin{equation*}
        \sum_{b\in \mZ_N}\sum_{j\in \mZ} \omega_N^{j\cdot( \langle \vec{a}_k, \vec{s}_0\rangle + b)}\cdot \rho_{r}(j) \Ket{b}.
	    \end{equation*}
	    Using the Poisson summation formula (Lemma~\ref{lem:PSF}) and changing the summation variable to $e \leftarrow N \cdot j + \langle \vec{a}_k, \vec{s}_0 \rangle+b$, the above state can be rewritten as
		 \begin{align*}
        \sum_{b\in \mZ_N}\sum_{j\in \mZ} \rho_{1/r}\Big(j + \frac{\langle \vec{a}_k, \vec{s}_0 \rangle + b}{N}\Big) \Ket{b}
        = 
        &\sum_{e \in \Z} \rho_{1/r}\Big(\frac{e}{N}\Big)  \Ket{\langle\vec{a}_k', \vec{s}_0\rangle + e \bmod N}  
        \end{align*}
        where $\vec{a}_k' = -\vec{a}_k \bmod N$. 
	    Since  $r= \Omega(\sqrt{\kappa})$,  we can apply Lemma~\ref{lem:gausstail_1} to the above state (for a scaled $\Z$-lattice), and instead of the above state,  consider the state that is within a $2^{-\WLandau(\kappa)}$ $\ell_2$-distance from it, namely:
        \begin{equation}\label{eq:LWEsuperp}
        \sum_{e\in \mZ_N}\rho_{1/r}\Big(\frac{e}{N}\Big) \Ket{\langle\vec{a}_k', \vec{s}_0\rangle + e}.
        \end{equation}
        Once we measure the state above, we obtain an LWE sample
        \[
        \left( \vec{a}_k', \langle\vec{a}_k', \vec{s}_0\rangle + e_k\right),
        \]
        where $e_k \hookleftarrow \mathcal{D}_{\mZ,N/(\sqrt{2}r)}$. We add to its second component an independent sample from $\mathcal{D}_{\mZ,N/(\sqrt{2}r)}$. By~\cite[Theorem~3.1]{Peikert10}, the resulting error is within statistical distance $2^{-\Omega(\kappa)}$ of $\mathcal{D}_{\mZ,N/r}$, as $N/(2r) > \sqrt{\kappa} \ge \eta_{\varepsilon}(\mZ)$ for some $\varepsilon = 2^{-\Omega(\kappa)}$.
        
Now we can call the $\LWE_{n,N,\alpha}$ oracle  for $\alpha = 1/r$ with the above states as input and obtain $\vec{s}_0$ as output of the oracle.
\end{proof}

\subsection{Reduction from $\DLMSS$ to $\DLWE$} \label{dLMSSdLWE}

Similar to the previous section where as a corollary we show that $\DLWE$ can be reduced to $\DLMSS$, we finish this section by a reverse reduction. Again we use exactly the same reduction algorithm as for the search versions (see Figure~\ref{fig:GLMSStoLWE}). Thus it remains to show that we can obtain a uniform random sample $(\vec{a}, b) \in \Z_N^n \times \Z_N$ given as input a state of the form $\Ket{j}\Ket{\vec{x} \bmod N}$.
\begin{corollary}[$\DLMSS \leq \DLWE$]\label{thm:dvDCPLWE}
	Let $(n, N, r)$ be valid $\DGLMSS$ parameters and $(n, N, \alpha)$ 
	be valid $\DLWE$ parameters, with $r = \Omega(\sqrt{\kappa})$ and $N > 2\sqrt{\kappa}\, r$. Given $\ell = \Omega(\kappa)$~many $\LMSS_{n,N,r}$ samples, there exists a quantum reduction, with run-time polynomial in $\kappa$, from $\DGLMSS_{n,N,r}^{\ell}$ to $\DLWE_{n,N,\alpha}^{\ell}$, where $\alpha = 1/r$.
\end{corollary}

\begin{proof}
	Assume we are given $\ell$~many samples of $\LMSS_{n,N,r}$ either of the form
	\[
	\left\{\sum_{j\in \mZ} \rho_r(j) \Ket{j}\Ket{\vec{x}_k+j\cdot \vec{s}_0 \bmod N}\right\}_{k\in [\ell]}
	\]
	or of the form
	\[
	\left\{\Ket{j_k}\Ket{\vec{x}_k \bmod N}\right\}_{k\in [\ell]},
	\]
	where $j_k \hookleftarrow \mathcal{D}^2_{\mZ,r}$ and $\vec{x}_k \in \mZ_N^n$ is uniform.
	Our aim is to distinguish between the above two forms given access to a $\DLWE_{n,N,\alpha}$ oracle.
	
	As explained above, we assume that random samples of $\LMSS$ are given. For each input state, after the quantum Fourier transform over $\mZ_N^n$ on the second register, we obtain
	\begin{equation*}
	\sum_{a\in \mZ_N^n} 
	\omega_N^{\langle \vec{x}_k,\vec{a}\rangle}
	\Ket{j_k} \Ket{\vec{a}}.
	\end{equation*}
	
	Then we measure the second register and let $\vec{a}_k$ denote the observed value. Note that each element of $\mZ_N^n$ is measured with 
	probability~$1/N^n$ and that the distributions for different $k$'s are independent. Up to a global phase, we have
	\begin{equation*}
	\Ket{j_k} \Ket{\vec{a}_k}.
	\end{equation*}
	We omit the second register which is known to us.
	According to Lemma~\ref{lem:gausstail_1} and since $N > 2\sqrt{\kappa}\, r$, with probability $1-2^{-\Omega(\kappa)}$, the value stored in the first register is in the range $[-\lfloor N/2 \rfloor, \lceil N/2 \rceil-1]$. Applying QFT over $\mZ_N$ to the first register, we obtain
	\begin{equation*}
	\sum_{b\in \mZ_N}\omega_N^{j_k\cdot b}\Ket{b}.
	\end{equation*}
	
	We measure the state above and let $b_k$ denote the observed value. Note that each element of $\mZ_N$ is measured with 
	probability~$1/N$ and that the distributions for different $k$'s are independent. We obtain a sample
	\[
	\left( \vec{a}_k, b_k\right),
	\]
	where $(\vec{a}_k, b_k)$ are uniformly random from $\Z_N^n \times \mZ_N$. This remains the case after adding to $b_k$ an independent sample from $\mathcal{D}_{\mZ,N/(\sqrt{2}r)}$, as in the proof of Theorem~\ref{thm:vDCPLWE2}.
\end{proof}


\bibliographystyle{plain}
\bibliography{bib/mybib}

@string{LNCS="LNCS"}

@string{Springer="Springer"}

@string{SIAM="SIAM"}

@string{Cambridge="Cambridge University Press"}

@string{ACM="ACM"}

@string{IEEE="IEEE Computer Society Press"}

@string{Combinatorica="Combinatorica"}

@string{MathAnn="Math.\ Ann"}

@string{JACM="J.\ {ACM}"}

@string{SIAMJC="{SIAM} J.\ Comput"}

@preamble{"\newcommand{\SortNoop}[1]{}"}

@article{HaWo94,
author = {P. Hausladen  and  W. K.   Wootters},
title = {A ‘Pretty Good’ Measurement for Distinguishing Quantum States},
journal = {Journal of Modern Optics},
volume = {41},
number = {12},
pages = {2385-2390},
year = {1994}
}

@article{BaChDa06,
  author    = {D. Bacon and A. M. Childs and
               W. van Dam},
  title     = {Optimal measurements for the dihedral hidden subgroup problem},
  journal   = {Chicago J. Theor. Comput. Sci.},
  year      = {2006}
}

@inproceedings{BaChDa05,
  author    = {D. Bacon and A. M. Childs and W. van Dam},
  title     = {From optimal measurement to efficient quantum algorithms for the hidden
               subgroup problem over semidirect product groups},
  booktitle = {Proc.\ or {FOCS}},
  pages     = {469--478},
  year      = {2005},
  publisher = IEEE
}

@inproceedings{BrVa11a,
  author    = {Brakerski, Z. and Vaikuntanathan, V.},
  title     = {Efficient Fully Homomorphic Encryption from (Standard) {LWE}},
  booktitle = {Proc.\ of {FOCS}},
  year      = {2011},
  pages     = {97--106},
  publisher = IEEE
}

@article{Peikert16,
  author    = {Peikert, C.},
  title     = {A Decade of Lattice Cryptography},
  journal   = {Foundations and Trends in Theoretical Computer Science},
  volume    = {10},
  number    = {4},
  pages     = {283--424},
  year      = {2016}
}

@article{Kuperberg05,
 author = {Kuperberg, G.},
 title = {A Subexponential-Time Quantum Algorithm for the Dihedral Hidden Subgroup Problem},
 journal = SIAMJC,
 issue_date = {2005},
 volume = {35},
 number = {1},
 year = {2005},
 pages = {170--188},
}

@inproceedings{EtHo99,
  author    = {M. Ettinger and P. H{\o}yer},
  title     = {On Quantum Algorithms for Noncommutative Hidden Subgroups},
  booktitle = {Proc.\ of {STACS}},
  pages     = {478--487},
  year      = {1999},
  series    = LNCS,
  volume    = {1563},
  publisher = Springer
}

@misc{GR02,
author = {Grover, L. and Rudolph, T.},
title = {Creating superpositions that correspond to efficiently integrable probability distributions},
note = {Draft. Available at \url{https://arxiv.org/pdf/quant-ph/0208112v1}},
year = 2002
}

@inproceedings{ChDa07,
  author    = {A. M. Childs and W. van Dam},
  title     = {Quantum algorithm for a generalized hidden shift problem},
  booktitle = {Proc.\ of {SODA}},
  pages     = {1225--1232},
  year      = {2007},
  publisher = SIAM
}

@inproceedings{Peikert10,
    author = {Peikert, C.},
    title = {An Efficient and Parallel {Gaussian} Sampler for Lattices},
    year = 2010,
  booktitle = {Proc.\ of {CRYPTO}},
  pages     = {80--97},
  publisher = Springer,
  series    = LNCS,
  volume    = {6223}
}

@inproceedings{SSTX09,
    Publisher = Springer,
    Title = {Efficient Public Key Encryption Based on Ideal Lattices},
    Series = LNCS,
    Booktitle = {Proc.\ of {ASIACRYPT}},
    Author = {Stehl{\'e}, D. and Steinfeld, R. and Tanaka, K. and Xagawa, K.},
    Volume = {5912},
    Year = 2009,
    Pages = {617--635}
}

@inproceedings{LyMi09,
    Author = {Lyubashevsky, V. and Micciancio, D.},
    Booktitle = {Proc.\ of {CRYPTO}},
    Title = {On Bounded Distance Decoding, Unique Shortest Vectors, and the Minimum Distance Problem},
    Year = 2009,
    Pages = {577-594}
}

@article{Banaszczyk93,
author={Banaszczyk, W.},
title={New bounds in some transference theorems in the geometry of numbers},
year={1993},
journal={Mathematische Annalen},
volume={296},
number={4},
publisher={Springer-Verlag},
pages={625-635},
}

@inproceedings{Regev02,
 author = {Regev, O.},
 title = {Quantum Computation and Lattice Problems},
 booktitle = {Proceedings of the 43rd Symposium on Foundations of Computer Science},
 series = {FOCS '02},
 year = {2002},
 pages = {520--529},
 numpages = {10},
 publisher = {IEEE Computer Society},
}

@article{Regev09,
  author    = {Regev, O.},
  title     = {On lattices, learning with errors, random linear codes, and cryptography},
  journal   = JACM,
  volume    = {56},
  number    = {6},
  year      = {2009}
}

@inproceedings{Peikert09,
    Publisher = ACM,
    Author = {Peikert, C.},
    Booktitle = {Proc.\ of {STOC}},
    Title = {Public-key cryptosystems from the worst-case shortest vector
               problem},
    Year = 2009,
    Pages = {333--342}
}

@book{BeBuDa08,
 author = {Bernstein, D. J. and Buchmann, J. and Dahmen, E.},
 title = {Post Quantum Cryptography},
 year = {2008},
 edition = {1st},
 publisher = Springer
}

@inproceedings{MiRe09,
    Publisher = Springer,
    Author = {Micciancio, D. and Regev, O.},
    Booktitle = {{Post-Quantum Cryptography},
                  D.~J.~Bernstein, J.~Buchmann, E.~Dahmen (Eds)},
    Title = {Lattice-based cryptography},
    Year = 2009,
    Pages = {147--191}
}

@inproceedings{GePeVa08,
    Publisher = ACM,
    Title = {Trapdoors for hard lattices and new cryptographic constructions},
    Booktitle = {Proc.\ of {STOC}},
    Author = {Gentry, C. and Peikert, C. and Vaikuntanathan, V.},
    Year = 2008,
    Pages = {197--206},
    note = {Full version available at \url{http://eprint.iacr.org/2007/432.pdf}}
}

@inproceedings{Regev05,
    Publisher = ACM,
    Title = {On lattices, learning with errors, random linear codes,
               and cryptography},
    Booktitle = {Proc.\ of {STOC}},
    Author = {Regev, O.},
    Year = 2005,
    Pages = {84--93}
}

@article{FIMSS14,
  author    = {K. Friedl and G. Ivanyos and F. Magniez and M. Santha and P. Sen},
  title     = {Hidden Translation and Translating Coset in Quantum Computing},
  journal   = SIAMJC,
  volume    = {43},
  number    = {1},
  pages     = {1--24},
  year      = {2014}
}

@article{Regev04,
  author    = {Regev, O.},
  title     = {New lattice-based cryptographic constructions},
  journal   = JACM,
  year      = {2004},
  volume    = {51},
  number    = {6},
  pages     = {899--942}
}

@article{Regev04b,
  author    = {O. Regev},
  title     = {Quantum Computation and Lattice Problems},
  journal   = SIAMJC,
  volume    = {33},
  number    = {3},
  pages     = {738--760},
  year      = {2004}
}

@article{Babai86,
    Title = {On {L}ov{\'a}sz lattice reduction and the nearest lattice point
problem},
    Journal = Combinatorica,
    Author = {Babai, L.},
    Volume = 6,
    Year = 1986,
    Pages = {1--13}
}

@article{Lenstra83,
    Title = {Integer programming with a fixed number of variables},
    Journal = {Mathematics of Operations Research},
    Author = {Lenstra, Jr., H.~W.},
    Number = 4,
    Volume = 8,
    Year = 1983,
    Pages = {538--548}
}

@inproceedings{GKPVZ13,
  author    = {Goldwasser, S. and Kalai, Y. T. and R. A. Popa and Vaikuntanathan, V. 
	and Zeldovich, N.},
  title     = {Reusable garbled circuits and succinct functional encryption},
  booktitle = {Proc.\ of {STOC}},
  pages     = {555--564},
  year      = {2013},
	publisher = ACM
}

@inproceedings{GVW13,
  author    = {Gorbunov, S. and Vaikuntanathan, V. and Wee, H.},
  title     = {Attribute-based encryption for circuits},
  booktitle = {Proc.\ of {STOC}},
  pages     = {545--554},
  year      = {2013},
  publisher = ACM
}

@article{LeLeLo82,
    Title = {Factoring polynomials with rational coefficients},
    Journal = MathAnn,
    Author = {Lenstra, A.~K. and Lenstra, Jr., H.~W. and Lov{\'a}sz, L.},
    Volume = 261,
    Year = 1982,
    Pages = {515--534}
}

@inproceedings{BLPRS13,
  Author = {Brakerski, Z. and Langlois, A. and Peikert, C. and Regev, O. and Stehl{\'e}, D.},
  title     = {Classical hardness of learning with errors},
  booktitle = {Proc.\ of {STOC}},
  year      = {2013},
  pages     = {575--584},
  publisher = ACM
}

@article{ORR13,
 author = {Ozols, M. and Roetteler, M. and Roland, J.},
 title = {Quantum Rejection Sampling},
 journal = {ACM Trans. Comput. Theory},
 issue_date = {August 2013},
 volume = {5},
 number = {3},
 month = aug,
 year = {2013},
 issn = {1942-3454},
 pages = {11:1--11:33},
 articleno = {11},
 numpages = {33},
 doi = {10.1145/2493252.2493256},
 acmid = {2493256},
 publisher = {ACM},
 address = {New York, NY, USA}
}

@book{Sho,
	author = {Shoup, Victor},
	title = {A Computational Introduction to Number Theory and Algebra},
	year = {2005},
	isbn = {0-5218-5154-8},
	publisher = {Cambridge University Press},
	address = {New York, NY, USA},
}

@misc{Fer26,
  author = {R. Fernandez},
  title = {Beyond {LWE}: Post-Quantum Assumptions under a Counterfactual Break of the Dihedral Coset Problem},
  year = 2026,
  howpublished = {arXiv:2610.36731},
}

\appendix

\section{Omitted Proofs from  Section~\ref{sec:pre}} \label{app:pre}

%
%

\subsection{Proof of Lemma~\ref{lem:gausstail_1}}\label{pro:gausstail_1}

\begin{proof}
	On one hand, from the Poisson summation formula (Lemma~\ref{lem:PSF}) for the lattice $\Z$, we have for $r > 0$
	\[
	\rho_{2r}(\Z) = \det(\Z) \cdot 2 r \cdot \sum_{y \in \Z} \rho_{1/(2r)}(y) \leq 2 r \cdot \sum_{y \in \Z} \rho_{1/r}(y) = 2 \rho_r(\Z),
	\]
	where for the inequality, we used the fact that $\rho_{1/(2r)} \leq \rho_{1/(r)}$.

	On the other hand,
	\begin{align*}
	\rho_{2r}(\Z) > \rho_{2r}(\Z \setminus [-B, B]) &= \sum_{\abs{y}>B} e^{-\frac{\pi y^2}{ (2r)^2}} \\
	& = \sum_{\abs{y}>B} e^{\frac{3}{4} \frac{\pi y^2}{(2r)^2} } e^{-\frac{\pi y^2}{r^2}} \geq  e^{\frac{3}{4} \frac{\pi (B+1)^2}{(2r)^2} } \rho_r(\Z \setminus [-B, B]).
	\end{align*}

	Combining the two inequalities for $\rho_{2r}$ and setting $B = \sqrt{\kappa}r$, we obtain 
	\[
	\rho_r(\Z \setminus [-\sqrt{\kappa}r, \sqrt{\kappa}r]) < e^{-\frac{3}{4} \frac{\pi (\sqrt{\kappa}r+1)^2}{(2r)^2} } \rho_{2r}(\Z) \le 2  e^{-\frac{3}{4} \frac{\pi (\sqrt{\kappa}r+1)^2}{(2r)^2} } \rho_r(\Z) = 2^{-\Omega(\kappa)} \rho_r(\mZ).
	\]
\end{proof}

%
%

\begin{revaddblock}
\subsection{Proof of Lemma~\ref{lem:infbound}}\label{pro:infbound-v3}

\begin{proof}
Let $Q = q^{(m-n)/m}$ and $B = Q/4$. If $Q < 4$, then $B < 1$ and the statement is trivial, as every nonzero vector of $\mZ^m$ has infinity norm at least~$1$. Assume now that $Q \ge 4$. Since $B < q$, a nonzero $\vec{y} \in \mZ^m$ with $\|\vec{y}\|_{\infty} < B$ is nonzero modulo~$q$; hence such a~$\vec{y}$ belongs to $\Lat_q(\vec{A})$ only if $\vec{y} \equiv \vec{A}\vec{x} \bmod q$ for some nonzero $\vec{x} \in \mZ_q^n$. For such an~$\vec{x}$, let $d = \gcd(x_1,\ldots,x_n,q)$, which is a proper divisor of~$q$, and write $\vec{x} = d\,\vec{x}'$ with $\vec{x}' \in \mZ_{q/d}^n$ satisfying $\gcd(x'_1,\ldots,x'_n,q/d) = 1$. Some integer combination of the coordinates of~$\vec{x}'$ is $1$ modulo~$q/d$, so that $\vec{A}\vec{x}' \bmod q/d$ is uniform in $\mZ_{q/d}^m$. As $\vec{A}\vec{x} \bmod q = d\cdot(\vec{A}\vec{x}' \bmod q/d)$, the probability $\Pr_{\vec{A}}[\vec{A}\vec{x} \equiv \vec{y} \bmod q]$ is $(q/d)^{-m}$ if $\vec{y} \in d\mZ^m$, and $0$ otherwise. For a given~$d$, there are at most $(q/d)^n$ such vectors~$\vec{x}$, and at most $(2B/d+1)^m$ vectors $\vec{y} \in d\mZ^m$ with $\|\vec{y}\|_{\infty} < B$; none of the latter is nonzero if $d \geq B$. By the union bound and since $q^{m-n} = Q^m$, we obtain
\begin{align*}
\Pr_{\vec{A}}\big[\lambda_1^{\infty}(\Lat_q(\vec{A})) < B\big] & \le \sum_{\substack{d \mid q,\ d<q \\ d < B}} \Big(\frac{q}{d}\Big)^{n-m} \Big(\frac{2B}{d}+1\Big)^m \\
& = \sum_{\substack{d \mid q,\ d<q \\ d < B}} \frac{1}{d^n} \Big(\frac{2B+d}{Q}\Big)^m \le \Big(\frac{3}{4}\Big)^{m} \sum_{\substack{d \mid q \\ d<q}} \frac{1}{d},
\end{align*}
where we used $2B+d \le 3B$ for $1 \le d \le B$. The last sum is at most $\sum_{k=1}^{q} 1/k \le 1+\ln q$.
\end{proof}
\end{revaddblock}

\begin{revaddblock}
\subsection{Proof of Lemma~\ref{lem:l2bound}}\label{pro:l2bound-v3}

\begin{proof}
Let $Q = q^{(m-n)/m}$ and $R = \min\{q, \sqrt{m}Q/(2\sqrt{2\pi e})\}$. Since $R \le q$, a nonzero $\vec{y} \in \mZ^m$ with $\|\vec{y}\| < R$ is nonzero modulo~$q$; hence such a~$\vec{y}$ belongs to $\Lat_q(\vec{A})$ only if $\vec{y} \equiv \vec{A}\vec{x} \bmod q$ for some nonzero $\vec{x} \in \mZ_q^n$. As in the proof of Lemma~\ref{lem:infbound}, for each proper divisor~$d$ of~$q$, there are at most $(q/d)^n$ vectors $\vec{x} \in \mZ_q^n$ with $\gcd(x_1,\ldots,x_n,q) = d$, and for each of them, the vector $\vec{A}\vec{x} \bmod q$ is uniform in $d\mZ_q^m$. Let $N(t)$ denote the number of nonzero $\vec{z} \in \mZ^m$ with $\|\vec{z}\| < t$. By the union bound, we obtain
\[
\Pr_{\vec{A}}\big[\lambda_1(\Lat_q(\vec{A})) < R\big] \le \sum_{d \mid q,\ d<q} T_d, \ \text{ with } \ T_d = \Big(\frac{q}{d}\Big)^{n-m} N\Big(\frac{R}{d}\Big).
\]
Fix~$d$ (possibly $d=1$), and let $q' = q/d$ and $Q' = q'^{(m-n)/m}$, so that $q'^{m-n} = Q'^m$. As $m \ge 2n$, we have $Q' \ge \sqrt{q'}$.

Assume first that $q' \ge 5$, so that $Q' \ge \sqrt{5}$. As $R \le \sqrt{m}Q/(2\sqrt{2\pi e})$ and $Q/d = Q' d^{-n/m} \le Q'$, we have $R/d \le u$ with $u = \sqrt{m}Q'/(2\sqrt{2\pi e})$. Let $\theta(s) = \sum_{k \in \mZ} e^{-\pi s k^2}$ for $s>0$. For every $s>0$, we have $N(u) \le \sum_{\vec{z} \in \mZ^m} e^{\pi s(u^2-\|\vec{z}\|^2)} = e^{\pi s u^2}\theta(s)^m$. We take $s = m/(2\pi u^2) = 4e/Q'^2$, and use the identity $\theta(s) = s^{-1/2}\theta(1/s)$, which follows from the Poisson summation formula (Lemma~\ref{lem:PSF}). We obtain $N(u) \le (Q'\,\theta(Q'^2/(4e))/2)^m$, and hence
\[
T_d \le Q'^{-m} N(u) \le \Big(\frac{\theta(Q'^2/(4e))}{2}\Big)^m \le \Big(\frac{3}{4}\Big)^m,
\]
as $\theta$ is decreasing and $\theta(5/(4e)) < 1.48$.

Assume now that $q' \le 4$. As $R \le q$, we have $R/d \le q' \le 4$. A nonzero $\vec{z} \in \mZ^m$ with $\|\vec{z}\| < 4$ satisfies $\|\vec{z}\|_1 \le \|\vec{z}\|^2 \le 15$, and is hence the sum of $15$ vectors from $\{\vec{0},\pm\vec{e}_1,\ldots,\pm\vec{e}_m\}$, repetitions allowed. This gives $N(R/d) \le (2m+1)^{15}$. As $q'^{n-m} \le 2^{-(m-n)} \le 2^{-m/2}$, we obtain $T_d \le (2m+1)^{15}\, 2^{-m/2}$. This case occurs for at most $3$ divisors~$d$, namely those with $q/d \in \{2,3,4\}$.

Summing over all proper divisors~$d$ of~$q$ gives the claim.
\end{proof}
\end{revaddblock}

\subsection{Proof of Lemma~\ref{lem:l2sforlmss}: $\LMSS$ self-reduction}\label{app:l2sforlmss}

\begin{proof}
We are given as input $\ell$ many $\GLMSS_{N,\ell,r_1}$ states:
\[
\left\{\sum_{j\in \mZ}\rho_{r_1}(j)\Ket{j}\Ket{(x_k+j\cdot s) 
\bmod N}\right\}_{k\le \ell}.
\]

Our aim is to find $s$, given access to a $\GLMSS_{N,\ell,r_2}$ oracle.

For each $\GLMSS_{N,\ell,r_1}$ sample, we proceed as follows. Using Lemma~\ref{lem:qrejsamp}, we can transform a $\GLMSS_{N,\ell,r_1}$ state into a $\GLMSS_{N,\ell,r_2}$ state with probability $\Omega(1/\kappa^c)$:
\[
\left\{\sum_{j\in \mZ}\rho_{r_2}(j)\Ket{j}\Ket{(x_k+j\cdot s) 
\bmod N}\right\}_{k\le \ell}.
\]

We repeat the above procedure for all given samples. With probability at least $\smash{ 1-2^{-\Omega(\kappa)}}$, we obtain $O(\ell/\kappa^{c+1})$ many $\GLMSS_{N,\ell,r_2}$ states. Then we can use the $\GLMSS_{N,O(\ell/\kappa^{c+1}),r_2}$ oracle to recover the secret $s$ as the solution for the $\GLMSS_{N,\ell,r_1}$ instance.

The same arguments apply to the uniform $\LMSS$: from $\ULMSS_{N,\ell,r_1}$ to $\ULMSS_{N,\bigO(\ell/\kappa^{c+1}),r_2}$. In this case, via quantum rejection sampling, we transform a wide uniform distribution into a narrow uniform distribution. 
\end{proof}

%
%


\section{Omitted Proofs from  Section~\ref{sec:LWEvDCP}} \label{app:LWEvDCP}

%
%

\subsection{Proof of Theorem~\ref{thm:LWEvDCP_ball}}\label{pro:LWEvDCP_ball}

We need two lemmata to establish a proof of Theorem~\ref{thm:LWEvDCP_ball}. The first lemma due to Regev (see \cite[Corollary 3.9]{Regev02}) shows how large the intersection of two shifted balls is. Note that a quantum state we work with is a discretized ball $\frac{1}{L}\mZ^n \cap \mathcal{B}_n(\vec{0},R)$. The difference between Regev's lemma and the one we present below is that we consider intersections of the balls taken $\bmod~q$.

\begin{lemma}[{Adapted from~\cite[Corollary 3.9]{Regev02}}]\label{lem:sphereintersect}
	Let $L = 2^n$ and consider the scaled integer grid $\frac{1}{L}\mZ^n$. For any $R \ge 1$, let $\bar{\mathcal{B}}_n(\vec{0},R) = \mathcal{B}_n(\vec{0},R) + \bar{d}$ for some vector $\bar{d}$ such that $R/\mathrm{poly}(n) \le \|\bar{d}\| \le R$. Then, for an integer $q>2R+\|\bar{d}\|$ we have
	\[
	\frac{|\frac{1}{L}\mZ^n \cap \mathcal{B}_n(\vec{0},R) \cap \bar{\mathcal{B}}_n(\vec{0},R) \bmod q|}{|\frac{1}{L}\mZ^n \cap \mathcal{B}_n(\vec{0},R) \bmod q|} \ge 1 - O(\sqrt{n}\|\bar{d}/R\|).
	\]
\end{lemma}

The following technical lemma is again due to Regev~\cite{Regev02}. It shows that while we cannot create a quantum state that perfectly corresponds to a uniform superposition of $x\in \frac{1}{L}\mZ^n \cap \mathcal{B}_n(\vec{0},R)$ (i.e., a ball), we can have a good approximation to this state. This fact will only increase the running time of the reduction by a factor of $\poly(\kappa)$.

\begin{lemma}[{\cite[Lemma 3.11]{Regev02}}]\label{lem:efficientstate}
	For any $1 \le R \le 2^{\mathrm{poly}(n)}$, let
	\[
	\Ket{\eta} = \frac{1}{\sqrt{|\frac{1}{L}\mZ^n \cap \mathcal{B}_n(\vec{0},R)|}} \sum_{x\in \frac{1}{L}\mZ^n \cap \mathcal{B}_n(\vec{0},R)}\Ket{x}
	\]
	be the uniform superposition on grid points inside a ball of radius $R$ around the origin where $L = 2^n$. Then, for any $c > 0$, a state $\Ket{\tilde{\eta}}$ whose trace distance from $\Ket{\eta}$ is at most $1/n^c$ can be efficiently computed.
\end{lemma}

Now we are ready to give the proof of Theorem~\ref{thm:LWEvDCP_ball}. 

\begin{proof}[of Theorem~\ref{thm:LWEvDCP_ball}]
Assume we are given an $\mathrm{LWE}_{n,q,\alpha}$ instance
$(\vec{A},\vec{b}_0) \in \mZ_q^{m\times n} \times \mZ_q^m$, with 
$\vec{b}_0=\vec{A} \vec{s}_0+\vec{e}_0 \bmod q$. 
Our aim is to find $\vec{s}_0$ given access to a $\GLMSS_{n,q,r}^{\ell}$ oracle.

We first prepare a necessary number of registers in the state $\Ket{0}$ and transform it to the state of the form (normalization omitted)
\begin{equation*}
\sum_{\vec{s} \in \mZ_q^n} \Ket{0} \Ket{\vec{s}} \Ket{0}.
\end{equation*}

We use Lemma~\ref{lem:gausssuperp}  to obtain a state 
within $\ell_2$ distance of $2^{-\Omega(\kappa)}$ away from 
\begin{equation*}
\sum_{\vec{s} \in \mZ_q^n} \mkern-7mu\Big(\sum_{j \in \mZ}\rho_{r}(j)\Ket{j}\Big)\Ket{\vec{s}}\Ket{0}.
\end{equation*}

According to Lemma~\ref{lem:gausstail_1}, the state above is within 
$\ell_2$ distance of $2^{-\Omega(\kappa)}$ away from
\[
\sum_{\substack{\vec{s} \in \mZ_q^n \\ j \in \mZ \cap [-\sqrt{\kappa} \cdot r,\sqrt{\kappa}\cdot r]}} \hspace*{-.9cm}
\rho_{r}(j)\Ket{j}
\Ket{\vec{s}}\Ket{0}.
\]

We evaluate the function $f(j,\vec{s}) \mapsto  \vec{A} \vec{s} - j\cdot \vec{b} \bmod q$ and store the result in the third register. A change of variable on $\vec{s}$ yields
\begin{equation*} 
    \sum_{\substack{ \vec{s} \in \mZ_q^n \\ j \in \mZ \cap [-\sqrt{\kappa} \cdot r,\sqrt{\kappa}\cdot r]}} \hspace*{-.8cm} 
		\rho_{r}(j)\Ket{j}\Ket{\vec{s}}\Ket{\vec{A} \vec{s}-j  \vec{A} \vec{s}_0-j \vec{e}_0} \ = \hspace*{-.8cm}
		\sum_{\substack{ \vec{s} \in \mZ_q^n \\ j \in \mZ \cap [-\sqrt{\kappa} \cdot r,\sqrt{\kappa}\cdot r]}} \hspace*{-.8cm} 
		\rho_{r}(j)\Ket{j}\Ket{\vec{s}+j \vec{s}_0} \Ket{\vec{A} \vec{s}-j \vec{e}_0}.
\end{equation*}

We take an auxiliary register $\Ket{0}$. 
We set $L = 2^m$ as a discretization parameter and $R = \TLandau(\sqrt{m}q^{(m-n)/m})$. By assumption on~$m$, we have $\sqrt{m}q^{(m-n)/m}/(3\sqrt{2\pi e}) \le q$. Thus according to~Lemma~\ref{lem:l2bound}, we also have $R \le \lambda_1(\Lambda_q(\vec{A}))/3$, where the inequality holds with probability at least $1-\tau(q)(3/4)^m-3(2m+1)^{15}2^{-m/2} = 1-2^{-\Omega(\kappa)}$, as the number $\tau(q)$ of divisors of~$q$ is at most~$q$. Further, the map $\vec{s} \mapsto \vec{A}\vec{s} \bmod q$ is injective on~$\mZ_q^n$ with probability at least $1-\sum_{d \mid q,\, d>1} d^{n-m} \ge 1-3\cdot 2^{n-m} = 1-2^{-\Omega(\kappa)}$ (see the proof of Theorem~\ref{thm:LWEvDCP}), so that distinct vectors~$\vec{s}$ give distinct lattice points~$\vec{A}\vec{s}$. Using Lemma~\ref{lem:efficientstate}, we prepare a state which is at most $1/n^c$ away from $\Ket{\eta} = \sum_{\vec{x} \in \frac{1}{L}\mZ^m \cap \mathcal{B}_m(\vec{0},R)} \Ket{\vec{x}}$ for any constant $c>0$. We set $c$ such that it satisfies $(1-1/n^c)>(1-1/\kappa)$ to have the success probability claimed at the end of the reduction. We pretend we work with the state $\Ket{\eta}$, which we tensor with our current state, and obtain 
\[
\sum_{\substack{ \vec{s} \in \mZ_q^n \\ j \in \mZ \cap [-\sqrt{\kappa} \cdot r,\sqrt{\kappa}\cdot r]}} 
\sum_{\vec{x} \in \frac{1}{L}\mZ^m \cap \mathcal{B}_m(\vec{0},R)}
\rho_{r}(j)\Ket{j}\Ket{\vec{s}+j\cdot \vec{s}_0}\Ket{\vec{A}\cdot \vec{s}-j\cdot \vec{e}_0}\Ket{\vec{x}}.
\]

We apply the function $g(\vec{x},\vec{y}) = (\vec{x},\vec{x}+\vec{y} \bmod q)$ on the third and fourth registers (note that the $\bmod~q$ operation is done over rational numbers). The state becomes a superposition over the balls of radius $R$ centered at all lattice points $\vec{A}\vec{s}$ and their shifts defined by $j$. Formally, we obtain
\begin{equation*} 
\sum_{\substack{ \vec{s} \in \mZ_q^n \\ j \in \mZ \cap [-\sqrt{\kappa} \cdot r,\sqrt{\kappa}\cdot r]}} 
\hspace*{-.5cm} 
\sum_{\vec{x} \in \frac{1}{L}\mZ^m \cap \mathcal{B}_{\text{\revdel{$n$}}\text{\revadd{$m$}}}(\vec{0},R)}
\hspace*{-.7cm}
\rho_{r}(j)\Ket{j}\Ket{\vec{s}+j\cdot \vec{s}_0}\Ket{\vec{A} \vec{s}-j\cdot \vec{e}_0}\Ket{\vec{A} \vec{s}-j\cdot \vec{e}_0 + \vec{x} \bmod q}. 
\end{equation*}

We measure the fourth register and discard it, as now we know it classically. 
Using Lemma~\ref{lem:shiftgaussiantail} on lattice $\mZ^m$ and shift $\vec{0}$, we have $\|\vec{e}_0\| \le \sqrt{m}\alpha q$ with probability $\ge 1-2^{-\Omega(m)} = 1-2^{-\Omega(\kappa)}$. Recall that  $r \le 1/(\sqrt{m\kappa}\ell\alpha q^{n/m})$ (up to constants). Therefore, we have $\|\sqrt{\kappa}r\cdot \vec{e}_0\| \le q^{(m-n)/m}/\ell$.

We use Lemma~\ref{lem:sphereintersect} with $\|\bar{d}\| \le 2q^{(m-n)/m}/\ell$ and radius  $R = \TLandau(\sqrt{m}q^{(m-n)/m})$, applied to the ball of each shift~$j$ and to each of the balls of the two extreme shifts $j' = \pm\lfloor\sqrt{\kappa}r\rfloor$, whose centers are at distance at most $2\|\sqrt{\kappa}r\cdot \vec{e}_0\|$ from its center. As balls are convex, a point lying in the balls of both extreme shifts lies in the balls of all shifts. Note that due to the minimal distance of $\qLat(\vec{A})$, the balls emerging from two different lattice points (or their shifts) do not intersect. Hence, the measurement yields
\[
\sum_{j \in \mZ \cap [-\sqrt{\kappa} \cdot r,\sqrt{\kappa}\cdot r]} 
\hspace*{-.3cm}
\rho_r(j)\Ket{j}\Ket{\vec{s}+j\cdot \vec{s}_0}\Ket{\vec{A} \vec{s}-j\cdot \vec{e}_0} 
\] 
for some $\vec{s} \in \mZ_q^n$, with probability $1-\bigO(1/\ell)$ taken over the uniform distribution defined on $\mathcal{B}_m(\vec{0},R)$.

Finally, we evaluate the function $(j,\vec{s},\vec{b}) \mapsto \vec{b}-\vec{A} \vec{s}+j\cdot\vec{b}_0$ on the first three registers. Discarding the last register, the state is of the form
\[
\sum_{j \in \mZ \cap [-\sqrt{\kappa} \cdot r,\sqrt{\kappa}\cdot r]}\hspace*{-.3cm}
\rho_{r}(j)\Ket{j}\Ket{\vec{s}+j\cdot \vec{s}_0}.
\]
According to Lemma~\ref{lem:gausstail_1}, the latter is within 
$\ell_2$ distance of $2^{-\Omega(\kappa)}$ away from
\begin{equation*} 
\sum_{j\in \mZ}\rho_{r}(j)\Ket{j}\Ket{\vec{s}+j\cdot \vec{s}_0}.
\end{equation*}

We repeat the above procedure $\ell$ times, and with non-negligible probability $(1-1/\ell)^{\ell}$, we obtain $\ell$ many $\GLMSS_{n,q,r}$ states
\[
\left\{\sum_{j\in \mZ}\rho_{r}(j)\Ket{j}\Ket{\vec{x}_k+j\cdot \vec{s}_0}\right\}_{k\le \ell},
\]
where $\vec{x}_k \in \mZ_q^n$.

Now we can call the $\GLMSS_{n,q,r}^{\ell}$ oracle with the above states as input and obtain $\vec{s}_0$ as the output of the oracle.
\end{proof}

\begin{proof}[of Corollary~\ref{cor:dLWEvDCP}]
\sloppy 
	We are given $m$~many samples from $\Z_q^{m \times n} \times \Z_q^m$ either of the form
	\[
	(\vec{A},\vec{A} \vec{s}_0+\vec{e}_0),
	\]
	or of the form
	\[
	(\vec{A},\vec{b}),
	\]
	where $\vec{A} \in \mZ_q^{m\times n}$ is uniformly chosen, $\vec{s}_0 \in \mZ_q^n$ is fixed, $\vec{e}_0 \hookleftarrow \mathcal{D}_{\mZ^m, \alpha q}$ and $\vec{b} \in \mZ_q^m$ is uniformly chosen and independent from $\vec{A}$.
	Our aim is to distinguish between the above two cases given access to a $\DGLMSS_{n,q,r}^{\ell}$ oracle.

	We follow the procedure of the reduction algorithm used in the proof of Theorem~\ref{thm:LWEvDCP_ball} except that now we set $R =\TLandau( \sqrt{m}q^{(m-n-1)/m})$. 
	By assumption on~$m$, we have $\sqrt{m}q^{(m-n-1)/m}/(3\sqrt{2\pi e}) \le q$. 
	Thus, in the case where $\vec{b}$ is uniform, we also have $R \leq \lambda_1(\Lambda_q(\vec{A}|\vec{b}))/3$, where the inequality holds with probability at least $1-\tau(q)(3/4)^m-3(2m+1)^{15}2^{-m/2} = 1-2^{-\Omega(\kappa)}$, as the number $\tau(q)$ of divisors of~$q$ is at most~$q$ (see Lemma~\ref{lem:l2bound}, applied to the uniform matrix $(\vec{A}|\vec{b})$). 
	Further, the map $\vec{x} \mapsto (\vec{A}|\vec{b})\vec{x} \bmod q$ is then injective on~$\mZ_q^{n+1}$ with probability at least $1-3\cdot 2^{n+1-m} = 1-2^{-\Omega(\kappa)}$ (see the proof of Theorem~\ref{thm:LWEvDCP}). In the case where $\vec{b} = \vec{A}\vec{s}_0+\vec{e}_0$, the lattice $\Lambda_q(\vec{A}|\vec{b})$ contains the short vector~$\vec{e}_0$, and we instead rely on the analysis of the proof of Theorem~\ref{thm:LWEvDCP_ball}, which applies with this smaller radius~$R$ thanks to the factor $q^{(n+1)/m}$ in the bound on~$r$. Thanks to the factor~$\varepsilon$ in the bound on~$r$, this analysis shows that each of the $\ell$ states is a $\GLMSS_{n,q,r}$ state with probability at least $1-\varepsilon/(4\ell)$, for $C$ large enough. Hence all $\ell$ states are valid except with probability at most~$\varepsilon/4$. Also, at the beginning of the algorithm, we prepare the ball states of Lemma~\ref{lem:efficientstate} with trace distance at most $\varepsilon/(8\ell)$ each, so $\ell$ of them contribute with at most~$\varepsilon/8$ distance from the correct distribution.
	
	We run the algorithm described in the proof of Theorem~\ref{thm:LWEvDCP_ball} until we obtain a state of the form
	\begin{equation*} \label{eq:StateU_gThm4}
	\sum_{\substack{ \vec{s} \in \mZ_q^n \\ j \in \mZ \cap [-\sqrt{\kappa} \cdot r,\sqrt{\kappa}\cdot r]}} 
	\sum_{\vec{x} \in \frac{1}{L}\mZ^m \cap \mathcal{B}_m(\vec{0},R)}
	\rho_{r}(j)\Ket{j}\Ket{\vec{s}}\Ket{\vec{A} \vec{s}-j\cdot \vec{b}}\Ket{\vec{A} \vec{s}-j\cdot \vec{b} + \vec{x}}.
	\end{equation*}
	
	We measure the fourth register and ignore it from now on. Due to the bound on $r$, we have $q \ge 2\sqrt{\kappa} r$; together with the injectivity of $\vec{x} \mapsto (\vec{A}|\vec{b})\vec{x} \bmod q$, this implies that distinct pairs $(\vec{s},j)$ give distinct points $\vec{A}\vec{s}-j\vec{b}$. Thus all the balls created with centers over the lattice points $\vec{A}\vec{s}$ and their shifts $\vec{A}\vec{s}-j\vec{b}$,
	do not intersect with each other. Therefore, we obtain
	\[
	\Ket{j_k}\Ket{\vec{s}_k}\Ket{\vec{A} \vec{s}_k-j_k \cdot \vec{b}},
	\] 
	for $j_k \in \mZ$ distributed as $\mathcal{D}^2_{\mZ,r}$ and some uniform $\vec{s}_k \in \mZ_q^n$.
	
	Finally, we evaluate the function $(j,\vec{s},\vec{b}) \mapsto \vec{b}-\vec{A} \vec{s}+j\cdot\vec{b}$ on the first three registers. 
	This zeroizes the last register, which we can safely discard. Now the state is of the form
	\[
	\Ket{j_k}\Ket{\vec{s}_k},
	\]
	where $j_k \hookleftarrow \mathcal{D}^2_{\mZ,r}$ and $\vec{s}_k \in \mZ_q^n$ is uniformly chosen.
	
	We repeat the above procedure $\ell$ times, and we obtain $\ell$ many random samples of the form
	\[
	\left\{\Ket{j_k}\Ket{\vec{s}_k}\right\}_{k\le \ell},
	\]
	where $\vec{s}_k$ is chosen uniformly at random from $\mZ_q^n$. This is exactly an input to  $\DLMSS$ with respect to Definition~\ref{def:DLMSS}. Therefore, the $\ell$ states we obtain are within statistical distance $\varepsilon/4+\varepsilon/8+2^{-\Omega(\kappa)}$ of $\ell$ $\GLMSS_{n,q,r}$ states with secret~$\vec{s}_0$ in the case of~$\vec{b}=\vec{A} \vec{s}_0+\vec{e}_0$, or are within statistical distance $\varepsilon/8+2^{-\Omega(\kappa)}$ of $\ell$ random samples as in Definition~\ref{def:DLMSS} in the case of~$\vec{b}$ uniform, respectively. Running the $\DGLMSS$ oracle on them therefore distinguishes the two cases with advantage at least $\varepsilon - 3\varepsilon/8 - \varepsilon/8 - 2^{-\Omega(\kappa)} \ge \varepsilon/2 - 2^{-\Omega(\kappa)}$.
\end{proof}

\end{document}